\documentclass[a4paper,10pt]{article}
\usepackage{jheppub}

\usepackage[utf8]{inputenc}
\usepackage{amsmath,amssymb,bm,slashed,braket}
\usepackage{dsfont}
\usepackage{amsfonts}
\usepackage{bm,bbm}
\usepackage{graphicx}
\usepackage{slashed}
\usepackage[dvipsnames,table]{xcolor}
\usepackage[normalem]{ulem}
\usepackage{soul}
\usepackage{siunitx} 
\usepackage{hyperref}
\hypersetup{colorlinks,citecolor= blue,linkcolor= blue, urlcolor=blue}
\usepackage{ulem}
\usepackage{array}
\usepackage{verbatim}
\usepackage{epsfig}
\usepackage{multirow, booktabs}
\usepackage{bbold}
\usepackage[version=4]{mhchem}
\usepackage[capitalise]{cleveref}
\usepackage{makecell}
\usepackage{environ}
\usepackage{amssymb}
\usepackage{pifont}

\allowdisplaybreaks[4]
\newcommand{\calO}{ {\cal O} }
\newcommand{\hc}{\text{H.c.}}

\usepackage{tikz} 
\usepackage{tkz-euclide}
\usetikzlibrary{backgrounds} 
\usepackage{circuitikz}
\usetikzlibrary{decorations.pathmorphing}
\usetikzlibrary{arrows.meta}
\tikzset{
mystyle/.style={line width=1, baseline, scale=0.6, every node/.style={scale=1}},
v/.style={decorate, draw, decoration={snake, segment length=2.mm, amplitude=0.5mm}},
f/.style={draw, decoration={markings,mark=at position #1 with {\arrow[]{Latex[length=1.5mm,width=1.5mm]}}},
    postaction={decorate},node contents=#1},
f/.default=.6,
fb/.style={draw,decoration={markings,mark=at position #1 with {\arrowreversed[]{Latex[length=1.5mm,width=1.5mm]}}},
    postaction={decorate},node contents=#1},
fb/.default=.6,
s/.style={dashed,draw, postaction={decorate},
        decoration={markings,mark=at position .55 with {\arrow[very thick]{latex}}}},
sb/.style={dashed,draw, postaction={decorate},
        decoration={markings,mark=at position .55 with {\arrowreversed[draw=black,very thick]{latex}}}},
snar/.style={dashed,draw,line width =1.25pt},
gluon/.style={decorate,
 decoration={coil,amplitude=2pt, segment length=3.5pt,  pre length=.1cm, post length=.1cm}},
}
\tikzset{mystyle/.style={line width=1,baseline,scale=0.6, every node/.style={scale=1}}}
\tikzset{circlestyle/.style={preaction={fill=white},postaction={pattern=north west lines},fill=cyan,fill opacity=1,draw=black}}
\tikzset{circlestyle2/.style={preaction={fill=white},postaction={pattern=north east lines},fill=red,fill opacity=0.5,draw=black}}
\tikzset{middlearrow/.style={
        decoration={markings,
            mark= at position 0.5 with {\arrow{#1}} ,
        },
        postaction={decorate}
    }
}

\title{Probing dimension-8 SMEFT operators through neutral meson mixing}

\author[a,b]{Yi Liao\mbox{\,\href{https://orcid.org/0000-0002-1009-5483}{\includegraphics[scale=0.075]{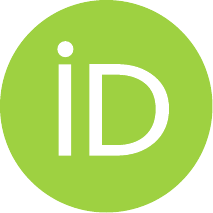}}}}
\emailAdd{liaoy@m.scnu.edu.cn}
\affiliation[a]{Key Laboratory of Atomic and Subatomic Structure and Quantum Control (MOE), 
Guangdong Basic Research Center of Excellence for Structure and Fundamental Interactions of Matter, 
Institute of Quantum Matter, South China Normal University, Guangzhou 510006, China}
\affiliation[b]{Guangdong-Hong Kong Joint Laboratory of Quantum Matter, 
Guangdong Provincial Key Laboratory of Nuclear Science, 
Southern Nuclear Science Computing Center, South China Normal University, Guangzhou 510006, China}
\author[a,b]{Xiao-Dong Ma\mbox{\,\href{https://orcid.org/0000-0001-7207-7793}{\includegraphics[scale=0.075]{Figs/orcid.pdf}}}}
\emailAdd{maxid@scnu.edu.cn}
\author[a,b]{Hao-Lin Wang\mbox{\,\href{https://orcid.org/0000-0002-2803-5657}{\includegraphics[scale=0.075]{Figs/orcid.pdf}}}}
\emailAdd{whaolin@m.scnu.edu.cn}

\abstract{
We investigate the impact of effective interactions of dimension-8 (dim-8) operators in the standard model effective field theory (SMEFT) on neutral meson mixing,
focusing on the $K^0-\bar K^0$, $B_{d,s}-\bar B_{d,s}$, and $D^0-\bar D^0$ systems. 
Within the framework of the low energy effective field theory (LEFT), 
each system is governed by eight dim-6 operators, 
with four originating at tree level from dim-6 SMEFT operators and the other four from dim-8 SMEFT operators. 
Notably, in certain UV complete models those dim-8 operators instead of the dim-6 ones are generated at the leading order. 
Our analysis focuses on those dim-8 operators and includes their one-loop QCD renormalization group running effects. 
By leveraging the LEFT master formula we impose stringent constraints on the effective scales associated with these dim-8 operators. 
We find that neutral meson mixing can probe an effective scale up to 80 TeV for some operators, 
surpassing the constraints imposed on other dim-8 operators by other observables. 
Lastly, we present a UV complete model capable of generating dim-8 operators at the leading order, 
thus offering a unique perspective on the interplay between different operator dimensions in probing new physics phenomena.
}

\keywords{Neutral meson mixing, standard model effective field theory, low energy effective field theory, renormalization group equations}

\makeatletter
\gdef\@fpheader{}
\makeatother
\begin{document}

\maketitle
\setcounter{page}{2}

\section{Introduction}

The rare flavor-changing neutral current (FCNC) processes are an important arena for probing new physics (NP) beyond the standard model (SM), primarily due to their significant suppression and well-controlled theoretical uncertainty in the SM prediction. Among these processes, the $\Delta F=1$, i.e., quark flavor changing by one unit, three-body semi-leptonic decays of mesons and baryons have demonstrated remarkable sensitivity to potential NP effects. In particular, there are long-standing anomalies in the lepton flavor universality observables like $R_{K^{(*)}}$ and $ R_{D^{(*)}}$ and potential NP implications from $K\to \pi \nu \bar \nu$ \cite{KOTO:2018dsc} and $B \to K \nu \bar \nu$ \cite{Belle-II:2023esi}. 

Furthermore, the $\Delta F=2$ processes such as neutral meson mixing ($M{\rm-}\bar M$ mixing)
offer a particularly fertile ground for exploring NP scenarios. The study of neutral meson mixing provides a rich source of information and insights into the dynamics of flavor physics and CP violation. By investigating the mixing phenomena we can glean valuable clues about the underlying mechanisms governing these processes. Many NP models that are related to neutrino mass generation,
dark matter interpretation, baryogenesis, and experimental anomalies can induce $\Delta F=2$ interactions, and the neutral meson mixing is a key observable to constrain the relevant parameter space.

Since the flavor changing processes mentioned above take place at a much lower energy below the SM electroweak scale $\Lambda_{\tt EW}$, they can be nicely described within the weak effective field theory (EFT) framework \cite{Buchalla:1995vs}. The weak EFT has been expanded in recent years into the more systematic low energy EFT (LEFT) \cite{Jenkins:2017jig,Liao:2020zyx}. On the other hand, the standard model effective field theory (SMEFT) above the electroweak scale is a powerful tool to parameterize indirect effects of heavy NP at the low energy scale \cite{Grzadkowski:2010es,Lehman:2014jma,Murphy:2020rsh,Liao:2016hru,Li:2020gnx,Liao:2020jmn,Li:2020xlh,Harlander:2023psl}. 
To study NP effects model-independently, one calculates the renormalization group (RG) running effect from the heavy NP scale ($\Lambda_{\tt NP}$) to $\Lambda_{\tt EW}$, 
performs the matching of the SMEFT interactions onto the LEFT at $\Lambda_{\tt EW}$, then calculates the RG running effect of the LEFT interactions from $\Lambda_{\tt EW}$ to the experimental scale, and finally formulates observables in terms of the SMEFT Wilson coefficients (WCs). 

The EFT approach has been systematically applied in previous studies of neutral meson mixing. 
For instance, Ref.\,\cite{Aebischer:2020dsw} provides a comprehensive analysis of dim-6 $\Delta F=2$ operators in SMEFT, including their matching to LEFT operators and leading order RG evolution effects. 
The next to leading order QCD running effects for these operators have been considered in \cite{Aebischer:2022anv}. Furthermore, Refs.\,\cite{ValeSilva:2022tph,Bhattacharya:2023beo} have studied the impacts on meson mixing of squared $\Delta F=1$ four-fermion operators in SMEFT. These studies are based on the general assumption that the NP induced FCNC interactions relevant to neutral meson mixing appear at the leading order as dim-6 operators in the SMEFT. But it may happen in some NP models that those dim-6 operators are highly suppressed or even not generated at the tree or even one-loop level due to symmetry \cite{He:2024iju,Liu:2016idz,Contino:2016jqw,Dawson:2022cmu}, so that the relevant operators first appear at the higher dimension 8. 
This is especially the case for models that include both vector-like heavy quarks and new scalars, which simultaneously couple to the SM chiral quarks through Yukawa interactions, as will be demonstrated later.
Considering severe suppression in SM of dim-6 operators we expect that the NP induced dim-8 operators may be stringently constrained in the absence of NP dim-6 operators. In this work we will fill this gap by investigating systematically the impact of those dim-8 operators on neutral meson mixing and employing the latest experimental measurements to constrain NP scenarios.

This paper is structured as follows. We start by presenting the relevant dim-8 operators in SMEFT and their matching onto the LEFT operators responsible for neutral meson mixing in \cref{sec:dim8}. Subsequently, we calculate the one-loop anomalous dimension matrix due to QCD interactions and solve the RG equations (RGEs) both analytically and numerically in Section \ref{sec:QCDrunning}. The detailed numerical analysis on neutral meson mixing due to the dim-8 SMEFT operators is given in \cref{sec4}, while in \cref{sec:UVmodel} we illustrate by a specific UV model that dim-8 instead of dim-6 operators arise at the leading order. Finally, we draw our conclusion in \cref{sec:conclusions}. In \cref{app:RGE} we provide the full one-loop QCD RGEs of the dim-8 operators in SMEFT. 
While in the main text we present our numerical results in the so-called up-quark flavor basis, we show our results in \cref{app:downbasis} in the down-quark flavor basis.
\section{The SMEFT dim-8 contribution to neutral meson mixing}
\label{sec:dim8}

LEFT is an effective field theory for the SM light quarks and leptons that respects the QCD and QED gauge symmetries $\rm SU(3)_c \times U(1)_{\rm em}$. In the LEFT framework, there are eight operators contributing to neutral meson mixing. The most commonly adopted basis is the SUSY basis \cite{Gabbiani:1996hi,FLAG:2019iem}, although other bases such as the BMU basis \cite{Buras:2000if} 
and the JMS basis \cite{Jenkins:2017jig} are also used in the literature (see, e.g., \cite{Buras:2022cyc}).
In the SUSY basis, they are parameterized as follows,  
\begin{subequations}
\label{eq:SUSY}
\begin{align}
\calO_{1}^{ij}&= (\bar{q}_{i}^\alpha \gamma_\mu P_L q_{j}^\alpha) (\bar{q}_{i}^\beta \gamma^\mu P_L q_{j}^\beta),
\\
\calO_{2}^{ij}&= (\bar{q}_{i}^\alpha P_L q_{j}^\alpha) (\bar{q}_{i}^\beta P_L q_{j}^\beta),
\\
\calO_{3}^{ij}&= (\bar{q}_{i}^\alpha P_L q_{j}^\beta) (\bar{q}_{i}^\beta P_L q_{j}^\alpha),
\\
\calO_{4}^{ij} &= (\bar{q}_{i}^\alpha P_L q_{j}^\alpha) (\bar{q}_{i}^\beta P_R q_{j}^\beta),
\\
\calO_{5}^{ij} &= (\bar{q}_{i}^\alpha P_L q_{j}^\beta) (\bar{q}_{i}^\beta P_R q_{j}^\alpha),
\\
\tilde\calO_{1,2,3}^{ij}&=\calO_{1,2,3}^{ij}|_{P_L\leftrightarrow P_R},
\end{align}
\end{subequations}
with $P_{R,L}=(1\pm\gamma_5)/2$ being the usual chiral projectors. Here the contracted Greek letters $\alpha$ and $\beta$ denote the color indices, while the flavor indices $ij=ds,\,cu,\,db,\,sb$ represent the involved operators responsible for $K^0-\bar K^0$, $D^0-\bar D^0$, $B_d-\bar B_d$, and $B_s-\bar B_s$ mixing, respectively. 
Note that the above convention implies $\tilde{\calO}_{2,3}^{ij}=\calO_{2,3}^{ji\,\dagger}$.

\begin{figure}
\centering
\begin{tikzpicture} [mystyle,scale=1.7, every node/.style={scale=0.7}]
\begin{scope}[shift={(1,1)}]
\filldraw[draw=green!30!white,fill=green!30!white] (0.,2.7) rectangle (11.6,3.4);
\node at (0.3,3.05){\bf D8: };
\node at (1,3.05){\footnotesize ${\cal O}_{\psi^4 H^2} $ };
\draw[thin,blue!80] (1,3.05) ellipse (0.3 and 0.3);
\node at (2,3.05){\footnotesize${\cal O}_{q^4H^2}^{(1)}$};
\draw[thin,blue!80] (2.5,3.05) ellipse (0.9 and 0.3);
\node at (3,3.05){\footnotesize${\cal O}_{q^4H^2}^{(3)}$};
\node at (4,3.05){\footnotesize${\cal O}_{q^2 \psi^2H^2}^{(1)}$};
\draw[thin,blue!80] (4.5,3.05) ellipse (0.9 and 0.3);
\node at (5,3.05){\footnotesize${\cal O}_{q^2 \psi^2H^2}^{(3)}$}; 
\node at (6,3.05){\footnotesize${\cal O}_{q^4 H^2}^{(2)}$};
\draw[thin,blue!80] (6.5,3.05) ellipse (0.9 and 0.3);
\draw[f,thick,orange] (6,2.7) -- (3.5,0);
\node at (7,3.05){\footnotesize${\cal O}_{q^4 H^2}^{(5)}$};
\node at (8,3.05){\footnotesize${\cal O}_{q^2\psi^2H^2}^{(2)}$};
\draw[thin,blue!80] (8.5,3.05) ellipse (0.9 and 0.3);
\draw[f,thick,orange] (8,2.7) -- (5.5,0);
\node at (9,3.05){\footnotesize${\cal O}_{q^2\psi^2H^2}^{(4)}$};
\draw[f,thick,orange] (9,2.7) -- (4.5,0);
\draw[f,thick,orange] (9,2.7) -- (5.5,0);
\node at (10,3.05){\footnotesize${\cal O}_{q^2\psi^2H^2}^{(5)}$};
\draw[thin,blue!80] (10.5,3.05) ellipse (0.9 and 0.3);
\draw[f,thick,purple] (10,2.7) -- (6.5,0);
\draw[f,thick,purple] (10,2.7) -- (7.5,0);
\node at (11,3.05){\footnotesize${\cal O}_{q^2\psi^2H^2}^{(6)}$};  
\draw[f,thick,purple] (11,2.7) -- (6.5,0);
\draw[f,thick,purple] (11,2.7) -- (7.5,0);
\draw[f,thick,purple] (11,2.7) -- (8.5,0);
\draw[f,thick,purple] (11,2.7) -- (9.5,0);
\filldraw[draw=gray!40!white,fill=gray!40!white] (0.,2.2) rectangle (5.4,2.7);
\node at (0.3,2.45){\bf D6: };
\node at (1,2.45){\scriptsize${\cal O}_{\psi\psi}$};
\draw[thin,blue!80] (1,2.45) ellipse (0.25 and 0.25);
\draw[f,thick] (1,2.2) -- (2.5,0);
\node at (2,2.45){\footnotesize${\cal O}_{qq}^{(1)}$};
\draw[thin,blue!80] (2.5,2.45) ellipse (0.8 and 0.25);
\draw[f,thick] (2,2.2) -- (3.5,0);
\node at (3,2.45){\footnotesize${\cal O}_{qq}^{(3)}$};
\draw[f,thick] (3,2.2) -- (3.5,0);
\node at (4,2.45){\footnotesize${\cal O}_{q \psi}^{(1)}$};
\draw[thin,blue!80] (4.5, 2.45) ellipse (0.8 and 0.25);
\draw[f,thick] (4,2.2) -- (5.5,0);
\node at (5,2.5){\footnotesize${\cal O}_{q \psi}^{(8)}$};
\draw[f,thick] (5,2.2) -- (4.5,0);
\draw[f,thick] (5,2.2) -- (5.5,0);
\draw[ultra thick,dashed] (0.65,3.4) rectangle (5.45,2.2);
\draw[draw=magenta!30,fill=magenta!30] (2.5,0) circle (0.22cm);
\node at (2.5,0){\footnotesize$\tilde{\cal O}_1 $};
\draw[draw=magenta!30,fill=magenta!30] (3.5,0) circle (0.22cm);
\node at (3.5,0){\footnotesize${\cal O}_1 $};
\draw[draw=magenta!30,fill=magenta!30] (4.5,0) circle (0.22cm);
\node at (4.5,0){\footnotesize${\cal O}_4 $};
\draw[draw=magenta!30,fill=magenta!30] (5.5,0) circle (0.22cm);
\node at (5.5,0){\footnotesize${\cal O}_5$};
\draw[draw=magenta!30,fill=magenta!30] (6.5,0) circle (0.22cm);
\node at (6.5,0){\footnotesize${\cal O}_2 $};
\draw[draw=magenta!30,fill=magenta!30] (7.5,0) circle (0.22cm);
\node at (7.5,0){\footnotesize$\tilde{\cal O}_2 $};
\draw[draw=magenta!30,fill=magenta!30] (8.5,0) circle (0.22cm);
\node at (8.5,0){\footnotesize${\cal O}_3 $};
\draw[draw=magenta!30,fill=magenta!30] (9.5,0) circle (0.22cm);
\node at (9.5,0){\footnotesize$\tilde{\cal O}_3 $};
\end{scope}
\end{tikzpicture}
\vspace{-0.8cm}
\caption{Matching correspondence between a dim-6 (D6) or dim-8 (D8) SMEFT operator 
and a dim-6 LEFT operator responsible for neutral meson mixing is indicated by an arrow line. 
Each light blue ellipse/circle encloses the set of operators that are mixed under QCD.}
\label{fig:flowdia}
\end{figure}

When working in the SMEFT framework, only four of the eight LEFT operators ($\calO_{1,4,5}$ and $\tilde \calO_1$) are generated from the dim-6 SMEFT operators via the tree-level matching, while the nontrivial contributions to the remaining four LEFT operators ($\calO_{2,3}$ and $\tilde \calO_{2,3}$) first appear at dimension 8 (see \cref{fig:flowdia} for the correspondence). The SMEFT dim-6 contributions to meson mixing including the RG effects have been considered extensively in the literature \cite{Aebischer:2020dsw,Aebischer:2022anv}, but a dim-8 analysis is still missing. Considering absence of dim-6 contributions at leading order in some NP models, it is important to include the contributions from dim-8 operators in the analysis. 
Since neutral meson mixing demands $\Delta F=2$, the relevant dim-8 SMEFT operators are those that contain an additional pair of Higgs fields on top of the four-fermion part. These operators can be classified into two types: the Type-I operators that are obtained by multiplying the corresponding dim-6 four-quark operators with the gauge invariant $H^\dagger H$, and the Type-II operators involving a nontrivial $\rm SU(2)_L$ structure of $H$. Together with dim-6 operators, all the dim-6 and dim-8 operators contributing to neutral meson mixing are collected in \cref{tab:SMEFTd8d6}. We adopt the Warsaw basis for the dim-6 operators \cite{Grzadkowski:2010es} and the notation from Ref.\,\cite{Murphy:2020rsh} for the dim-8 operators. In the table, the SM left-handed quark doublet is denoted by $q$, while the right-handed up- and down-type quark singlets are denoted by $u$ and $d$, respectively.

\begin{table}
\center
\resizebox{\linewidth}{!}{
\renewcommand\arraystretch{1.9}
\begin{tabular}{| l | c | l | l | c | l | l | c |}
\hline
\multicolumn{2}{|c}{\bf{Dim 6}} & \multicolumn{6}{|c|}{\bf{Dim 8}} 
\\
\hline
\multicolumn{2}{|c}{-}  & \multicolumn{2}{|c|}{\bf{Type-I}} & $\#$ & \multicolumn{2}{c|}{\bf{Type-II}} & $\#$ 
\\
\hline
$\calO^{(1)}_{qq}$ & $(\bar q \gamma_\mu q) (\bar q \gamma^\mu q)$ 
& $\calO_{q^4 H^2}^{(1)}$ & $\calO_{qq}^{(1)} (H^\dagger H)$ & $45$
& $\calO_{q^4 H^2}^{(2)}$  & $(\overline{q}\gamma^\mu q) (\overline{q}\gamma_\mu \tau^I q) (H^\dagger \tau^I H)$ & $81$
\\
$\calO_{qq}^{(3)}$ & $(\bar q \gamma_\mu \tau^I q) (\bar q \gamma^\mu \tau^I q)$  
 & $\calO_{q^4 H^2}^{(3)}$ & $\calO_{qq}^{(3)} (H^\dagger H)$ & $45$
 & $\calO_{q^4 H^2}^{(5)}$  & $i \epsilon^{IJK} (\overline{q}\gamma^\mu \tau^I q) (\overline{q}\gamma_\mu \tau^J q) (H^\dagger \tau^K H)$ & $36$ 
\\
\hline
$\calO^{(1)}_{qu}$ & $(\bar q \gamma_\mu q) (\bar u\gamma^\mu u)$ 
& $\calO_{q^2 u^2 H^2}^{(1)}$ & $\calO_{qu}^{(1)} (H^\dagger H)$ & $81$  
& $\calO_{q^2 u^2 H^2}^{(2)}$ & $(\overline{q}\gamma^\mu \tau^I q) (\overline{u}\gamma_\mu u) (H^\dagger \tau^I H)$ & $81$ 
\\
$\calO^{(8)}_{qu}$ & $(\bar q \gamma_\mu T^A q) (\bar u\gamma^\mu T^A u)$   
& $\calO_{q^2 u^2 H^2}^{(3)}$ & $\calO_{qu}^{(8)}(H^\dagger H)$ & $81$ 
& $\calO_{q^2 u^2 H^2}^{(4)}$ & $(\overline{q}\gamma^\mu T^A \tau^I q) (\overline{u}\gamma_\mu T^A u) (H^\dagger \tau^I H)$ & $81$
\\
\hline
$\calO^{(1)}_{qd}$ & $(\bar q \gamma_\mu q) (\bar d\gamma^\mu d)$ 
& $\calO_{q^2 d^2 H^2}^{(1)}$ & $\calO_{qd}^{(1)} (H^\dagger H)$ & $81$
& $\calO_{q^2 d^2 H^2}^{(2)}$ & $(\overline{q}\gamma^\mu \tau^I q) (\overline{d}\gamma_\mu d) (H^\dagger \tau^I H)$ & $81$
\\
$\calO^{(8)}_{qd}$ & $(\bar q \gamma_\mu T^A q) (\bar d\gamma^\mu T^A d)$ 
  & $\calO_{q^2 d^2 H^2}^{(3)}$ & $\calO_{qd}^{(8)} (H^\dagger H)$ & $81$
  & $\calO_{q^2 d^2 H^2}^{(4)}$ & $(\overline{q}\gamma^\mu T^A \tau^I q) (\overline{d}\gamma_\mu T^A d) (H^\dagger \tau^I H)$ & $81$
\\
\hline
$\calO_{uu}$ & $(\bar{u}\gamma_\mu u) (\bar{u}\gamma^\mu u)$ 
& $\calO_{u^4 H^2}$ & $\calO_{uu} (H^\dagger H)$ & $45$ 
& $\calO_{q^2 u^2 H^2}^{(5)}\,(+\hc)$ & $(\overline{q} u \Tilde{H}) (\overline{q} u \Tilde{H})$ & $45+45$ 
\\
$\calO_{dd}$ & $(\bar{d}\gamma_\mu d) (\bar{d}\gamma^\mu d)$ 
& $\calO_{d^4 H^2}$ & $\calO_{dd} (H^\dagger H)$ & $45$  
& $\calO_{q^2 u^2 H^2}^{(6)}\,(+\hc)$ & $(\overline{q} T^A u \Tilde{H}) (\overline{q} T^A  u \Tilde{H})$ & $45+45$ 
\\
\hline
 &  & & &  &$\calO_{q^2 d^2 H^2}^{(5)}\,(+\hc)$ & $(\overline{q} d H) (\overline{q} d H)$ & $45+45$
\\
 &  & & & & $\calO_{q^2 d^2 H^2}^{(6)}\,(+\hc)$ & $(\overline{q} T^A d H) (\overline{q} T^A  d H)$ & $45+45$ 
\\
\hline
\end{tabular} 
}
\caption{The leading dim-6 and dim-8 SMEFT operators that contribute to neutral meson mixing.
We include an imaginary unit $i$ for the operator $\calO_{q^4 H^2}^{(5)}$, 
and $+\hc$ denotes Hermitian conjugate of the preceding effective interaction.
} 
\label{tab:SMEFTd8d6}
\end{table}

\begin{table}[htb]
\center
\resizebox{\linewidth}{!}{
\renewcommand\arraystretch{1.9}
\begin{tabular}{| c | l |}
\hline
\cellcolor{magenta!25}LEFT operators & \multicolumn{1}{|c|}{\cellcolor{magenta!25}Matching results at electroweak scale $\Lambda_{\tt EW}$}
\\
\hline
\multirow{2}*{$\calO_{1}^{ij}= (\bar{q}_{i}^\alpha \gamma_\mu P_L q_{j}^\alpha) (\bar{q}_{i}^\beta \gamma^\mu P_L q_{j}^\beta)$}
 & $C_{1,dd}^{ij}=
 \left[ { v^2\over 2}  
 \left( C_{q^4 H^2}^{(2),xyzw} 
 {\color{cyan} + C_{q^4 H^2}^{(1),xyzw} + C_{q^4 H^2}^{(3),xyzw} }
 \right) 
{\color{blue} + C_{qq}^{(1),xyzw} + C_{qq}^{(3),xyzw} }
 \right] V_{xi}^* V_{yj} V^*_{zi} V_{wj}$
\\
\cline{2-2}
& $C_{1,uu}^{ij}= -{ v^2\over 2} \left(  C_{q^4 H^2}^{(2),ijij} {\color{cyan} - C_{q^4 H^2}^{(1),ijij} - C_{q^4 H^2}^{(3),ijij} } \right)
{\color{blue} + C_{qq}^{(1),ijij} + C_{qq}^{(3),ijij}}$
\\
\hline
\multirow{2}*{$\tilde \calO_{1}^{ij}= (\bar{q}_{i}^\alpha \gamma_\mu P_R q_{j}^\alpha) (\bar{q}_{i}^\beta \gamma^\mu P_R q_{j}^\beta)$}
 & $\tilde C_{1,dd}^{ij}= {\color{cyan}{v^2 \over 2} C^{ijij}_{d^4 H^2}} {\color{blue} + C_{dd}^{ijij}}$
\\
\cline{2-2}
& $\tilde C_{1,uu}^{ij}= {\color{cyan}{v^2 \over 2} C^{ijij}_{u^4 H^2}}{\color{blue} + C_{uu}^{ijij}}$
\\
\hline
\multirow{2}*{$\calO_{2}^{ij}= (\bar{q}_{i}^\alpha P_L q_{j}^\alpha) (\bar{q}_{i}^\beta P_L q_{j}^\beta)$} 
& $C^{ij}_{2,dd}= {v^2 \over 2}\left(C^{(5),xizi*}_{q^2 d^2 H^2}-{1\over 6}C^{(6),xizi*}_{q^2 d^2 H^2}  \right)V_{xj} V_{zj}  $
\\
\cline{2-2}
 & $C^{ij}_{2,uu}= {v^2 \over 2} 
 \left( C^{(5),jiji*}_{q^2 u^2 H^2} -{1 \over 6} C^{(6),jiji*}_{q^2 u^2 H^2} \right)$
\\
\hline
\multirow{2}*{$\calO_{3}^{ij} = (\bar{q}_{i}^\alpha P_L q_{j}^\beta) (\bar{q}_{i}^\beta P_L q_{j}^\alpha)$} 
& $C_{3,dd}^{ij} = {v^2 \over 4} C^{(6),xizi *}_{q^2 d^2 H^2} V_{xj} V_{zj}$
\\
\cline{2-2}
 & $C_{3,uu}^{ij} = {v^2 \over 4} C^{(6),jiji *}_{q^2 u^2 H^2}$
 \\
\hline
\multirow{2}*{$\calO_{4}^{ij} = (\bar{q}_{i}^\alpha P_L q_{j}^\alpha) (\bar{q}_{i}^\beta P_R q_{j}^\beta)$} 
& $C_{4,dd}^{ij} = 
\left[ 
- {v^2 \over 2} \left( C^{(4),xyij}_{q^2 d^2 H^2} 
{\color{cyan} + C_{q^2 d^2 H^2}^{(3),xyij} }  \right) 
{\color{blue}-C_{qd}^{(8),xyij}} \right]
V_{xi}^* V_{yj}$
\\
\cline{2-2}
 & $C_{4,uu}^{ij}={v^2 \over 2} \left( 
 C^{(4),ijij}_{q^2 u^2 H^2} {\color{cyan} - C^{(3),ijij}_{q^2 u^2 H^2} }
 \right)
 {\color{blue}-C_{qu}^{(8),ijij}}$
\\
\hline
\multirow{2}*{$\calO_{5}^{ij} = (\bar{q}_{i}^\alpha P_L q_{j}^\beta) (\bar{q}_{i}^\beta P_R q_{j}^\alpha)$} & 
$C_{5,dd}^{ij} = 
\left[ - v^2 \left( 
C^{(2),xyij}_{q^2 d^2 H^2} - {1\over 6} C^{(4),xyij}_{q^2 d^2 H^2}  
{\color{cyan}
+C_{q^2 d^2 H^2}^{(1),xyij} - {1\over 6 } C_{q^2 d^2 H^2}^{(3),xyij} } \right) 
{\color{blue}- 2 C_{qd}^{(1),xyij} + {1\over 3 }C_{qd}^{(8),xyij}}
\right] V_{xi}^* V_{yj}$
\\
\cline{2-2}
 & $C_{5,uu}^{ij}=v^2 \left( C^{(2),ijij}_{q^2 u^2 H^2}
 -{1\over 6} C^{(4),ijij}_{q^2 u^2 H^2} {\color{cyan}- C_{q^2 u^2 H^2}^{(1),ijij}+{1\over 6} C^{(3),ijij}_{q^2 u^2 H^2} } \right) 
 {\color{blue} -2 C_{qu}^{(1),ijij} + {1\over3}C_{qu}^{(8),ijij}}$
\\
\hline
\end{tabular} 
}
\caption{The $\Delta F=2$ tree-level matching results from relevant dim-6 (in blue) and dim-8 operators in the SMEFT. The terms in cyan (black) denote the contributions from type-I (type-II) dim-8 operators.}
\label{tab:matching}
\end{table}

The column $\#$ in \cref{tab:SMEFTd8d6} counts the total number of independent operators after taking into account three quark flavors.  
Notice that the Type-II operator $\calO_{q^4 H^2}^{(5)}$ is anti-symmetric under the interchange of the flavor indices in the two quark currents, i.e., 
$\calO_{q^4 H^2}^{(5),xyzw}= - \calO_{q^4 H^2}^{(5),zwxy}$, while the Type-II operators 
$\calO_{q^2 \psi^2 H^2}^{(5,6)}$ and Type-I operators $\calO_{q^4 H^2}^{(1,3)}, \calO_{\psi^4 H^2}$ 
(together with the corresponding dim-6 operators) are symmetric under the same operation, where $\psi=u,d$. 
The flavor symmetry of these operators leads to the same flavor symmetry structure in their corresponding WCs. 
The contribution to the meson mixing from other SMEFT dim-8 operators with four quark fields like 
$(\bar q q)^2H D$ or  $(\bar q q)^2D^2, (\bar q q)^2G$ (where $D$ stands for the covariant derivative 
and $G$ for the gluon field strength tensor) is suppressed relative to that from those in \cref{tab:SMEFTd8d6} 
by a factor of $\Lambda_{\tt QCD}/v \sim 10^{-3}$ or $\Lambda_{\tt QCD}^2/v^2 \sim 10^{-6}$, and thus is neglected. 
Here $\Lambda_{\tt QCD}\sim 200\,\rm MeV$ is the typical nonperturbative QCD scale and $v\sim 246\,\rm GeV$ the Higgs vacuum expectation value (vev).

Next we turn to the tree-level matching of the above SMEFT and LEFT interactions. Due to the family $\rm SU(3)$ symmetry in the SM quark sector, we have the freedom to choose either up-type or down-type quarks as the mass eigenstates, but we will do numerical analyses in both bases as in Ref.\,\cite{Aebischer:2020dsw}: the up-type basis in this section and the down-type basis in \cref{app:downbasis}. 
We assume that the up-type quarks are already in their mass eigenstates and that the diagonalization matrix for right-handed down-type quarks has been incorporated into the definition of WCs. Then, quark flavor mixing enters manifestly only through the Cabibbo-Kobayashi-Maskawa (CKM) matrix acting on left-handed down-type quarks, $d_{Lx}'= V_{x i} d_{Li}$, where $x$ ($i$) refers to the family (flavor) of quarks, for instance, $V_{1s}=V_{us}$. Upon sending $H\to v/\sqrt{2}$, we match the SMEFT onto the LEFT at $\Lambda_{\tt EW}\equiv v/\sqrt{2}\approx 160\,\rm GeV$. The matching relations for the WCs of the  LEFT $\Delta F=2$ interactions are summarized in \cref{tab:matching}; for tree-level matching in the JMS basis \cite{Jenkins:2017jig}, see also Ref.\,\cite{Hamoudou:2022tdn}.
In the table, we show both the dim-6 and dim-8 contributions to the $\Delta F =2$ LEFT WCs relevant to neutral meson mixing. As mentioned above, the matching results for the WCs of the other two operators $\tilde \calO_{2,3}^{ij}$ are simply the identification: $\tilde C_{2,3}^{ij} = C_{2,3}^{ji*}$.
Thus, the NP contribution to the effective Hamiltonian describing the $\Delta F=2$ meson mixing at 
the scale $\mu=\Lambda_{\tt EW}$ becomes, 
\begin{align}
\label{eq:effH}
- {\cal H}_{\Delta F=2}^{{\tt NP},ij}(\Lambda_{\tt EW})
= \sum_{a=1}^5 C_a^{ij} \calO_a^{ij} 
+\sum_{b=1}^3 \tilde C_b^{ij} \tilde \calO_b^{ij}, 
\end{align}
where we have suppressed the scale dependence on $\Lambda_{\tt EW}$ on the right-hand side.

The above matching provides the direct contribution to meson mixing from local $\Delta F=2$ dim-6 and dim-8 SMEFT operators. Note that double insertions of $\Delta F=1$ dim-6 SMEFT operators in the classes $\{ \psi^2H^3, \psi^2 H^2 D, \psi^2 X H, \psi^4\}$ can also contribute at the same order of power counting as dim-8 operators. The $\Delta F=1$ operators containing a single quark current in the classes $\{ \psi^2H^3, \psi^2 H^2 D, \psi^2 X H\}$ contribute to meson mixing through tree-level diagrams involving mediators such as $h/Z/\gamma$, while $\Delta F=1$ four-fermion operators in the class $\psi^4$ contribute via one-loop diagrams. Among these, the double insertions of $\Delta F=1$ operators in the classes $\psi^2H^3$ and $\psi^4$ have been studied in Ref.\,\cite{Harnik:2012pb} and Refs.\,\cite{ValeSilva:2022tph,Bhattacharya:2023beo}, respectively. Since the $\Delta F=1$ operators in the class $\psi^2 X H$ involve a field strength tensor that can only be generated at a loop level from UV models, their contribution to meson mixing from double insertions is expected to be highly suppressed due to the presence of double loop factors. 
Last, the $\Delta F=1$ operators in the class $\psi^2 H^2 D$ also contribute to the $\Delta F=1$ FCNC meson ($M$) decays $M_1\to M_2 + \ell^+ \ell^-/\bar\nu\nu$ and $M \to \ell^+ \ell^-/\bar\nu\nu$ via a $Z$ mediator, and are thus strongly constrained by experimental measurements. In summary, if any of the above dim-6 operators could be induced from a UV model and its  double insertion could contribute 
equally importantly as the dim-8 $\Delta F=2$ operators, it should be included in the discussion of meson mixing. Nevertheless, we will suggest a UV model in \cref{sec:UVmodel} that cannot induce such dim-6 operators at the leading order.

\section{The QCD RG effect for dim-8 SMEFT operators}
\label{sec:QCDrunning}

\begin{figure}[t]
	\centering
	\begin{tikzpicture}[mystyle,scale=1.1]
		\begin{scope}[shift={(1,1)}]
			\draw[f,blue] (0,0) -- (-1.5,1.5);
			\draw[f, blue] (1.5,1.5)-- (0,0);
            \draw[f,blue] (0,0) -- (-1.5,-1.5);
            \draw[f,blue] (1.5,-1.5) -- (0,0);
			\draw[gluon, magenta, thick] (-1.0,1.0) -- (1.0,1.0);
            \draw[snar, thick] (1.5,0) -- (0,0);
            \draw[snar, thick] (0,0) -- (-1.5,0);
			\draw[draw=cyan,fill=cyan] (0,0) circle (0.2cm);
            \draw[thin, blue] (-0.25, 0.55) arc (190:350:0.25) 
            node[currarrow,pos=0.5, xscale=-1,sloped,scale=0.6,blue] {};
            \draw[thin, blue] (0.25,-0.55) arc (10:170:0.25)
            node[currarrow,pos=0.5, xscale=-1,sloped,scale=0.6,blue] {};
            \node[yshift=-1.2cm] at (0,0) {$(a)$};
		\end{scope}
	\end{tikzpicture} \qquad\quad
    \begin{tikzpicture}[mystyle,scale=1.1]
		\begin{scope}[shift={(1,1)}]
			\draw[f,blue] (0,0) -- (-1.5,1.5);
			\draw[f, blue] (1.5,1.5)-- (0,0);
            \draw[f,blue] (0,0) -- (-1.5,-1.5);
            \draw[f,blue] (1.5,-1.5) -- (0,0);
			\draw[gluon, magenta, thick] (-1.0,1.0) -- (-1.0,-1.0);
            \draw[snar, thick] (1.5,0) -- (0,0);
            \draw[snar, thick] (0,0) -- (-1.5,0);
			\draw[draw=cyan,fill=cyan] (0,0) circle (0.2cm);
            \draw[thin, blue] (-0.25, 0.55) arc (190:350:0.25) 
            node[currarrow,pos=0.5, xscale=-1,sloped,scale=0.6,blue] {};
            \draw[thin, blue] (0.25,-0.55) arc (10:170:0.25)
            node[currarrow,pos=0.5, xscale=-1,sloped,scale=0.6,blue] {};
            \node[yshift=-1.2cm] at (0,0) {$(b)$};
		\end{scope}
	\end{tikzpicture}\qquad\quad
    \begin{tikzpicture}[mystyle,scale=1.1]
		\begin{scope}[shift={(1,1)}]
			\draw[f,blue] (0,0) -- (-1.5,1.5);
			\draw[f, blue] (1.5,1.5)-- (0,0);
            \draw[f,blue] (0,0) -- (-1.5,-1.5);
            \draw[f,blue] (1.5,-1.5) -- (0,0);
            \draw [gluon,magenta, thick] (-1.0,1.0) arc (135:-45:1.414);
            \draw[snar, thick] (1.5,0) -- (0,0);
            \draw[snar, thick] (0,0) -- (-1.5,0);
			\draw[draw=cyan,fill=cyan] (0,0) circle (0.2cm);
            \draw[thin, blue] (-0.25, 0.55) arc (190:350:0.25) 
            node[currarrow,pos=0.5, xscale=-1,sloped,scale=0.6,blue] {};
            \draw[thin, blue] (0.25,-0.55) arc (10:170:0.25)
            node[currarrow,pos=0.5, xscale=-1,sloped,scale=0.6,blue] {};
            \node[yshift=-1.2cm] at (0,0) {$(c)$};
		\end{scope}
	\end{tikzpicture}
\qquad\quad
 \begin{tikzpicture}[mystyle,scale=1.1]
		\begin{scope}[shift={(1,1)}]
			\draw[f,blue] (0,0.7) -- (-1.5,1.5);
			\draw[f, blue] (1.5,1.5)-- (0,0.7);
            \draw[thick,blue] (0,0.7) arc (-270:90:0.55);
            \draw[gluon,magenta, thick] (0,-0.4) -- (0,-1.4);
            \draw[snar, thick] (1.5,0.7) -- (0,0.7);
            \draw[snar, thick] (0,0.7) -- (-1.5,0.7);
			\draw[draw=cyan,fill=cyan] (0,0.7) circle (0.2cm);
            \draw[thin, blue] (-0.25, 1.3) arc (190:350:0.25) 
            node[currarrow,pos=0.5, xscale=-1,sloped,scale=0.6,blue] {};
            \draw[thin, blue] (0.25,0.2) arc (10:170:0.25)
            node[currarrow,pos=0.5, xscale=-1,sloped,scale=0.6,blue] {};
            \node[yshift=-1.2cm] at (0,0) {$(g)$};
		\end{scope}
  	\end{tikzpicture}
    \\
    \vspace{0.2cm}
    \begin{tikzpicture}[mystyle,scale=1.1]
	   \begin{scope}[shift={(1,1)}]
			\draw[f,blue] (0,0) -- (-1.5,1.5);
			\draw[f, blue] (1.5,1.5)-- (0,0);
            \draw[f,blue] (0,0) -- (-1.5,-1.5);
            \draw[f,blue] (1.5,-1.5) -- (0,0);
			\draw[gluon, magenta, thick] (-1.0,-1.0) -- (1.0,-1.0);
            \draw[snar, thick] (1.5,0) -- (0,0);
            \draw[snar, thick] (0,0) -- (-1.5,0);
			\draw[draw=cyan,fill=cyan] (0,0) circle (0.2cm);
            \draw[thin, blue] (-0.25, 0.55) arc (190:350:0.25) 
            node[currarrow,pos=0.5, xscale=-1,sloped,scale=0.6,blue] {};
            \draw[thin, blue] (0.25,-0.55) arc (10:170:0.25)
            node[currarrow,pos=0.5, xscale=-1,sloped,scale=0.6,blue] {};
            \node[yshift=-1.2cm] at (0,0) {$(d)$};
		\end{scope}
	\end{tikzpicture} \qquad\quad
 \begin{tikzpicture}[mystyle,scale=1.1]
		\begin{scope}[shift={(1,1)}]
			\draw[f,blue] (0,0) -- (-1.5,1.5);
			\draw[f, blue] (1.5,1.5)-- (0,0);
            \draw[f,blue] (0,0) -- (-1.5,-1.5);
            \draw[f,blue] (1.5,-1.5) -- (0,0);
			\draw[gluon, magenta, thick] (1.0,1.0) -- (1.0,-1.0);
            \draw[snar, thick] (1.5,0) -- (0,0);
            \draw[snar, thick] (0,0) -- (-1.5,0);
			\draw[draw=cyan,fill=cyan] (0,0) circle (0.2cm);
            \draw[thin, blue] (-0.25, 0.55) arc (190:350:0.25) 
            node[currarrow,pos=0.5, xscale=-1,sloped,scale=0.6,blue] {};
            \draw[thin, blue] (0.25,-0.55) arc (10:170:0.25)
            node[currarrow,pos=0.5, xscale=-1,sloped,scale=0.6,blue] {};
            \node[yshift=-1.2cm] at (0,0) {$(e)$};
		\end{scope}
	\end{tikzpicture}\qquad\quad
    \begin{tikzpicture}[mystyle,scale=1.1]
		\begin{scope}[shift={(1,1)}]
			\draw[f,blue] (0,0) -- (-1.5,1.5);
			\draw[f, blue] (1.5,1.5)-- (0,0);
            \draw[f,blue] (0,0) -- (-1.5,-1.5);
            \draw[f,blue] (1.5,-1.5) -- (0,0);
            \draw[gluon,magenta, thick] (-1.0,-1.0) arc (225:45:1.414);
            \draw[snar, thick] (1.5,0) -- (0,0);
            \draw[snar, thick] (0,0) -- (-1.5,0);
			\draw[draw=cyan,fill=cyan] (0,0) circle (0.2cm);
            \draw[thin, blue] (-0.25, 0.55) arc (190:350:0.25) 
            node[currarrow,pos=0.5, xscale=-1,sloped,scale=0.6,blue] {};
            \draw[thin, blue] (0.25,-0.55) arc (10:170:0.25)
            node[currarrow,pos=0.5, xscale=-1,sloped,scale=0.6,blue] {};
            \node[yshift=-1.2cm] at (0,0) {$(f)$};
		\end{scope}
	\end{tikzpicture}\qquad\quad
 \begin{tikzpicture}[mystyle,scale=1.1]
		\begin{scope}[shift={(1,1)}]
			\draw[f,blue] (0,0.7) -- (-1.5,1.5);
			\draw[f, blue] (1.5,1.5)-- (0,0.7);
            \draw[thick,blue] (0,0.7) arc (-270:90:0.55);
            \draw[gluon,magenta, thick] (0,-0.4) -- (0,-1.4);
            \draw[snar, thick] (1.5,0.7) -- (0,0.7);
            \draw[snar, thick] (0,0.7) -- (-1.5,0.7);
			\draw[draw=cyan,fill=cyan] (0,0.7) circle (0.2cm);
            \draw[thin, blue] (0.7, 0.45) arc (-100:-260:0.25) 
            node[currarrow,pos=0.5, xscale=-1,sloped,scale=0.6,blue] {};
            \draw[thin, blue] (-0.7,0.95) arc (80:-80:0.25)
            node[currarrow,pos=0.5, xscale=-1,sloped,scale=0.6,blue] {};
            \node[yshift=-1.2cm] at (0,0) {$(h)$};
		\end{scope}
	\end{tikzpicture}
\caption{One-loop Feynman diagrams for the irreducible QCD corrections. 
The cyan blobs represent the insertion of dim-8 interactions,
and the blue solid (black dashed, magenta wiggly) lines denote quark (Higgs, gluon) fields. }
	\label{fig:QCDoneloop}
\end{figure}

Since the SMEFT operators are usually generated at a much higher scale than $\Lambda_{\tt EW}$,
their contributions to low energy observables are affected by the RG effect, with the leading one coming from QCD interactions. 
Following the method in our previous works \cite{Liao:2019tep,Liao:2019gex}, we calculate one-loop diagrams shown in \cref{fig:QCDoneloop} with the insertion of each dim-8 operator $\calO_A^B$ together with its corresponding WC $C_A^B$. These diagrams are grouped into two classes: with an internal gluon propagator ($a,b,c,d,e,f$) or with an external gluon leg ($g,h$). The former one is renormalized by the same set of dim-8 operators, while the latter one induces the redundant gluon equation of motion (EoM) operators, $(\bar q \gamma^\mu T^A \tau^I q) D^\nu G^{A}_{\mu\nu} (H^\dagger \tau^I H)$ and $(\bar \Psi \gamma^\mu T^A \Psi) D^\nu G^{A}_{\mu\nu} (H^\dagger H)$ with $\Psi=q,u,d$. 
By employing the EoMs for the gluons, $D^\mu G^{A}_{\mu,\nu} = g_s \sum_{\Psi=q,u,d}\bar \Psi \gamma_\nu T^A \Psi$, those operators reduce to four-quark-two-Higgs operators but with $\Delta F=1$ and are thus not relevant to the $\Delta F=2$ meson mixing.
For the purpose of reference, the full RGEs including the gluon EoM contributions are summarized in \cref{app:RGE}. Besides the one-particle-irreducible diagrams shown in \cref{fig:QCDoneloop}, one also has to take into account the wavefunction renormalization. After obtaining all these one-loop QCD corrections, the RGEs  can be extracted from their pole  terms.

Since the Type-I and Type-II operators have different $\rm SU(2)_L$ structures for the quark parts, there is no mixing between the two types of operators under the QCD renormalization. For the Type-II dim-8 operators, we obtain the following RGEs of the WCs, 
\begin{subequations}
\label{eq:RGEtypeII}
    \begin{align}
        \mu\frac{d}{d\mu}
        \begin{pmatrix}
		C^{(2),xyzw}_{q^4 H^2}
		\\
		C^{(2),xwzy}_{q^4 H^2}
        \\
        C^{(2),zyxw}_{q^4 H^2}
        \\
        C^{(2),zwxy}_{q^4 H^2}
        \\
        C^{(5),xyzw}_{q^4 H^2}
        \\
        C^{(5),xwzy}_{q^4 H^2}
	   \end{pmatrix}
	   &= -\frac{\alpha_s}{2\pi}
	   \begin{pmatrix}
		{3 \over N_c} & -{3\over 2} & -{3\over 2} & 0 & 0 & -6 \\
		-{3\over 2} & {3 \over N_c} & 0 & -{3\over 2} & -6 & 0 \\
		-{3\over 2} & 0 & {3 \over N_c} & -{3\over 2} & 6 & 0 \\
        0 & -{3\over 2} & -{3\over 2} & {3 \over N_c} & 0 & 6 \\
		0 & -{3\over 4} & {3\over 4} & 0 & {3 \over N_c} & 0 \\
		-{3\over 2} & 0 & 0 & {3\over 4} & 0 & {3 \over N_c} 
	   \end{pmatrix}
	   \begin{pmatrix}
		C^{(2),xyzw}_{q^4 H^2}
		\\
		C^{(2),xwzy}_{q^4 H^2}
        \\
        C^{(2),zyxw}_{q^4 H^2}
        \\
        C^{(2),zwxy}_{q^4 H^2}
        \\
        C^{(5),xyzw}_{q^4 H^2}
        \\
        C^{(5),xwzy}_{q^4 H^2}
	   \end{pmatrix},
        \\
        \mu\frac{d}{d\mu}
        \begin{pmatrix}
		C^{(2),xyzw}_{q^2 \psi^2 H^2}
		\\
		C^{(4),xyzw}_{q^2 \psi^2 H^2}
	   \end{pmatrix}
	   &= -\frac{\alpha_s}{2\pi}
	   \begin{pmatrix}
		0 & {3\over N_c} C_F  \\
		6 & 6 C_F - {3 \over N_c} 
	   \end{pmatrix}
	   \begin{pmatrix}
		C^{(2),xyzw}_{q^2 \psi^2 H^2}
		\\
		C^{(4),xyzw}_{q^2 \psi^2 H^2}
	   \end{pmatrix},
        \\
        \mu\frac{d}{d\mu}
        \begin{pmatrix}
		C^{(5),xyzw}_{q^2 \psi^2 H^2}
		\\
		C^{(5),xwzy}_{q^2 \psi^2 H^2}
        \\
		C^{(6),xyzw}_{q^2 \psi^2 H^2}
        \\
		C^{(6),xwzy}_{q^2 \psi^2 H^2}
	   \end{pmatrix}
	   &= -\frac{\alpha_s}{2\pi}
	   \begin{pmatrix}
6C_F & { 4- 4N_c^2\over N_c^2} & { N_c^2- 4C_F N_c -1 \over N_c^2 } & {2(N_c^2-C_F  N_c^3-1)\over N_c^3} \\
{ 4- 4N_c^2\over N_c^2} & 6C_F & {2(N_c^2-C_F N_c^3-1)\over N_c^3} &  { N_c^2- 4C_F N_c -1 \over N_c^2 } \\
-4 & {8\over N_c} & -2 C_F & -{4\over N_c^2}-2 \\
{8\over N_c} & - 4 & -{4\over N_c^2}-2 & -2 C_F
	   \end{pmatrix}
	    \begin{pmatrix}
		C^{(5),xyzw}_{q^2 \psi^2 H^2}
		\\
		C^{(5),xwzy}_{q^2 \psi^2 H^2}
        \\
		C^{(6),xyzw}_{q^2 \psi^2 H^2}
        \\
		C^{(6),xwzy}_{q^2 \psi^2 H^2}
	   \end{pmatrix},
    \end{align}
\end{subequations}
where the subscript label $\psi = u, d$ and $\alpha_s = g_s^2/(4\pi)$. 
$N_c =3$ is the number of colors and $C_F =4/3$ the Casimir invariant of $\rm SU(3)_{\rm c}$. 
Due to flavor mixing, we introduce the repeated WCs with flavor reshuffled and organize the results in the matrix form 
that will facilitate solving the differential equations in an analytical fashion. The RGEs for the Type-I operators are,
\begin{subequations}
\label{eq:RGEtypeI}
    \begin{align}
         \mu\frac{d}{d\mu}
        \begin{pmatrix}
		C^{(1),xyzw}_{q^4 H^2}
		\\
		C^{(1),xwzy}_{q^4 H^2}
        \\
        C^{(3),xyzw}_{q^4 H^2}
        \\
		C^{(3),xwzy}_{q^4 H^2}
	   \end{pmatrix}
	   &= -\frac{\alpha_s}{2\pi}
	   \begin{pmatrix}
		{3\over N_c} & -{3\over 2} & 0 & -{9\over 2} \\
		-{3\over 2} & {3\over N_c} & -{9\over 2} & 0 \\
        0 & -{3\over 2} & {3\over N_c} & {3\over 2} \\
        -{3\over 2} & 0 & {3\over 2} & {3\over N_c}
	   \end{pmatrix}
	   \begin{pmatrix}
		C^{(1),xyzw}_{q^4 H^2}
		\\
		C^{(1),xwzy}_{q^4 H^2}
        \\
        C^{(3),xyzw}_{q^4 H^2}
        \\
		C^{(3),xwzy}_{q^4 H^2}
	   \end{pmatrix},
    \\
    \mu\frac{d}{d\mu}
        \begin{pmatrix}
		C^{(1),xyzw}_{q^2 \psi^2 H^2}
		\\
		C^{(3),xyzw}_{q^2 \psi^2 H^2}
	   \end{pmatrix}
	   &= -\frac{\alpha_s}{2\pi}
	   \begin{pmatrix}
		0 & {3\over N_c} C_F \\
		6 &   6 C_F-{3\over N_c}
	   \end{pmatrix}
	   \begin{pmatrix}
		C^{(1),xyzw}_{q^2 \psi^2 H^2}
		\\
		C^{(3),xyzw}_{q^2 \psi^2 H^2}
	   \end{pmatrix},
    \\
    \mu\frac{d}{d\mu}
      \begin{pmatrix}
		C^{xyzw}_{\psi^4 H^2}
		\\
		C^{xwzy}_{\psi^4 H^2}
	   \end{pmatrix}
	   &= -\frac{\alpha_s}{2\pi}
	   \begin{pmatrix}
		{3\over N_c} & -3 \\
		-3 & {3\over N_c}
	   \end{pmatrix}
	   \begin{pmatrix}
		C^{xyzw}_{\psi^4 H^2}
		\\
		C^{xwzy}_{\psi^4 H^2}
	   \end{pmatrix}.  
    \end{align}
\end{subequations}

To solve the above RGEs, we use the one-loop QCD $\beta$ function, $\mu{d\alpha_s \over d\mu} = {b_0 \over 2\pi} \alpha_s^2 $, 
where $b_0= -11 + 2n_f/3$ with $n_f$ being the number of active quarks between the scales $\mu_1$ and $\mu_2$. 
We denote the diagonal matrices formed with the eigenvalues of the anomalous dimension matrices in the above equations as:
\begin{subequations}
\begin{align}
 R[C_{q^4 H^2}^{(2,5)}] &= {\rm diag}\left(\zeta_{2/1}^{4 \over b_0},\zeta_{2/1}^{4 \over b_0},\zeta_{2/1}^{4 \over b_0} ,\zeta_{2/1}^{-{ 2\over b_0} },\zeta_{2/1}^{- { 2\over b_0}},\zeta_{2/1}^{- { 2\over b_0}} \right), 
 \\
R[C_{q^2 \psi^2 H^2}^{(2,4)}] &= {\rm diag}\left(\zeta_{2/1}^{8 \over b_0},\zeta_{2/1}^{- { 1\over b_0}} \right), 
\\
R[C_{q^2 \psi^2 H^2}^{(5,6)}] &= {\rm diag}\left(\zeta_{2/1}^{17 + \sqrt{241} \over 3 b_0},
\zeta_{2/1}^{- { \sqrt{241} + 1 \over 3 b_0} },
\zeta_{2/1}^{ {\sqrt{241} - 1 \over 3 b_0} },
\zeta_{2/1}^{ {17 - \sqrt{241} \over 3 b_0} }\right), 
\\
R[C_{q^4 H^2}^{(1,3)}] &= {\rm diag}\left(\zeta_{2/1}^{4 \over b_0},\zeta_{2/1}^{4 \over b_0},\zeta_{2/1}^{- { 2\over b_0}},\zeta_{2/1}^{- { 2\over b_0}} \right), 
 \\
 R[C_{\psi^4 H^2}] &= {\rm diag}\left(\zeta_{2/1}^{4 \over b_0},\zeta_{2/1}^{- { 2\over b_0}} \right), 
\end{align}
\end{subequations}
where $\zeta_{2/1}=\alpha_s(\mu_2)/\alpha_1(\mu_1)$ and
$C_{q^2 \psi^2 H^2}^{(1,3)}$ has the same RGE as $C_{q^2 \psi^2 H^2}^{(2,4)}$, 
$R[C_{q^2 \psi^2 H^2}^{(1,3)}] = R[C_{q^2 \psi^2 H^2}^{(2,4)}] $.
The corresponding diagonalization matrices are
\begin{subequations}
\begin{align}
T[C_{q^4 H^2}^{(2,5)}]  & = 
\left(
\begin{array}{cccccc}
 -\frac{1}{8} & 0 & 0 & \frac{1}{8} & 0 & \frac{1}{2} \\
 0 & -\frac{1}{8} & \frac{1}{8} & 0 & \frac{1}{2} & 0 \\
 0 & -\frac{1}{4} & -\frac{1}{4} & \frac{1}{2} & 0 & 1 \\
 \frac{1}{8} & 0 & 0 & -\frac{1}{8} & 0 & \frac{1}{2} \\
 0 & \frac{1}{8} & -\frac{1}{8} & 0 & \frac{1}{2} & 0 \\
 0 & \frac{1}{4} & \frac{1}{4} & \frac{1}{2} & 0 & -1 \\
\end{array}
\right), 
\\
T[C_{q^2 \psi^2 H^2}^{(2,4)}] & = \left(
\begin{array}{cc}
 \frac{2}{3} & \frac{8}{9} \\
 -\frac{2}{3} & \frac{1}{9} \\
\end{array}
\right), 
\\
T[C_{q^2 \psi^2 H^2}^{(5,6)}] & = 
\left(
\begin{array}{cccc}
 \frac{5}{\sqrt{241}} & -\frac{5}{\sqrt{241}} & -\frac{1}{4} + \frac{53}{12 \sqrt{241}} & \frac{1}{4}-\frac{53}{12 \sqrt{241}} \\
 \frac{1}{\sqrt{241}} & \frac{1}{\sqrt{241}} & \frac{1}{4}+\frac{43}{12 \sqrt{241}} & \frac{1}{4}+\frac{43}{12 \sqrt{241}} \\
 -\frac{1}{\sqrt{241}} & -\frac{1}{\sqrt{241}} & \frac{1}{4}-\frac{43}{12 \sqrt{241}} & \frac{1}{4}-\frac{43}{12 \sqrt{241}} \\
 -\frac{5}{\sqrt{241}} & \frac{5}{\sqrt{241}} & -\frac{1}{4}-\frac{53}{12 \sqrt{241}} & \frac{1}{4}+\frac{53}{12 \sqrt{241}} \\
\end{array}
\right), 
\\
T[C_{q^4 H^2}^{(1,3)}]  & = 
\left(
\begin{array}{cccc}
 -\frac{1}{4} & 0 & \frac{1}{4} & \frac{1}{2} \\
 0 & -\frac{1}{4} & \frac{1}{2} & \frac{1}{4} \\
 \frac{1}{4} & 0 & -\frac{1}{4} & \frac{1}{2} \\
 0 & \frac{1}{4} & \frac{1}{2} & -\frac{1}{4} \\
\end{array}
\right), 
\\
T[C_{\psi^4 H^2}]  & = \left(
\begin{array}{cc}
 -\frac{1}{2} & \frac{1}{2} \\
 \frac{1}{2} & \frac{1}{2} \\
\end{array}
\right),
\end{align}    
\end{subequations}
and $T[C_{q^2 \psi^2 H^2}^{(1,3)}] = T[C_{q^2 \psi^2 H^2}^{(2,4)}]$. 
Note that each diagonalization matrix is unique up to a  multiplication constant. 
With the above results, the solution to a WC column vector $C_x$ appearing in \cref{eq:RGEtypeII} or \cref{eq:RGEtypeI} between the two scales $\mu_1$ and $\mu_2$ is
\begin{align}
C_x(\mu_1) = T[C_x] \, R[C_x] \, T[C_x]^{-1} \, C_x(\mu_2).     
\end{align}
By choosing $\mu_2 = 5 \,\rm TeV$ and $\mu_1 = \Lambda_{\tt EW}$, for the Type-II operators we find numerically, 
\begin{subequations}
\label{eq:RGRsolution}
\begin{align}
C_{q^4 H^2}^{(2),xyzw}(\Lambda_{\tt EW}) & = 1.06 C_{q^4 H^2}^{(2),xyzw}   
- 0.08 (C_{q^4 H^2}^{(2),xwzy} + C_{q^4 H^2}^{(2),zyxw} ) 
- 0.31 C_{q^4 H^2}^{(5),xwzy}, 
\\
C_{q^4 H^2}^{(5),xyzw}(\Lambda_{\tt EW}) & = 
- 0.04 ( C_{q^4 H^2}^{(2),xwzy} - C_{q^4 H^2}^{(2),zyxw}) 
+ 1.06 C_{q^4 H^2}^{(5),xyzw},
\\
C_{q^2 \psi^2 H^2}^{(2), xyzw}(\Lambda_{\tt EW}) & = 
1.01  C_{q^2 \psi^2 H^2}^{(2), xyzw}
+ 0.08 C_{q^2 \psi^2 H^2}^{(4), xyzw}, 
\\
C_{q^2 \psi^2 H^2}^{(4), xyzw}(\Lambda_{\tt EW}) & = 
0.36  C_{q^2 \psi^2 H^2}^{(2), xyzw}
+ 1.43 C_{q^2 \psi^2 H^2}^{(4), xyzw}, 
\\
C_{q^2 \psi^2 H^2}^{(5),xyzw}(\Lambda_{\tt EW}) & = 1.51 C_{q^2 \psi^2 H^2}^{(5),xyzw}
- 0.25 C_{q^2 \psi^2 H^2}^{(5),xwzy}
- 0.03 C_{q^2 \psi^2 H^2}^{(6),xyzw}
- 0.11 C_{q^2 \psi^2 H^2}^{(6),xwzy},
\\
C_{q^2 \psi^2 H^2}^{(6),xyzw}(\Lambda_{\tt EW}) & =
-0.25 C_{q^2 \psi^2 H^2}^{(5),xyzw}
+ 0.19 C_{q^2 \psi^2 H^2}^{(5),xwzy}
+ 0.88 C_{q^2 \psi^2 H^2}^{(6),xyzw}
- 0.10 C_{q^2 \psi^2 H^2}^{(6),xwzy},
\end{align}
\end{subequations} 
and for the Type-I operators the results are
\begin{subequations} 
\begin{align}
C_{q^4 H^2}^{(1),xyzw}(\Lambda_{\tt EW}) & = 
1.06 C_{q^4 H^2}^{(1),xyzw}
-0.08 C_{q^4 H^2}^{(1),xwzy}
-0.24 C_{q^4 H^2}^{(3),xwzy},
\\
C_{q^4 H^2}^{(3),xyzw}(\Lambda_{\tt EW}) & = 
-0.08 C_{q^4 H^2}^{(1),xwzy}
+1.06 C_{q^4 H^2}^{(3),xyzw}
+0.08 C_{q^4 H^2}^{(3),xwzy}, 
\\
C_{q^2 \psi^2 H^2}^{(1),xyzw}(\Lambda_{\tt EW}) & = 
1.01  C_{q^2 \psi^2 H^2}^{(1), xyzw}
+ 0.08 C_{q^2 \psi^2 H^2}^{(3),xyzw}, 
\\
C_{q^2 \psi^2 H^2}^{(3),xyzw}(\Lambda_{\tt EW}) & = 
0.36 C_{q^2 \psi^2 H^2}^{(1),xyzw}
+ 1.43 C_{q^2 \psi^2 H^2}^{(3),xyzw}, 
\\
C_{\psi^4 H^2}^{xyzw}(\Lambda_{\tt EW}) & = 
1.06 C_{\psi^4 H^2}^{xyzw} 
- 0.16 C_{\psi^4 H^2}^{xwzy},
\end{align}
\end{subequations} %
where the scale $\mu_2 = 5\,\rm TeV$ on the right-hand side is not shown explicitly for brevity.
From the above results we notice that, even though the operator $\calO^{(5)}_{q^4 H^2}$ has no direct contribution to the matching results in \cref{tab:matching}, 
the mixing effect through RG running would make it relevant. Nevertheless, due to the anti-symmetric property of its WC under flavor interchange, 
$C_{q^4 H^2}^{(5),xyzw}=-C_{q^4 H^2}^{(5),zwxy}$, it still has no contribution to the flavor symmetric LEFT operator $\calO_1^{ij}$ as shown in \cref{tab:matching}.

\section{The LEFT master formula and the constraints}
\label{sec4}

%

In the previous sections, we have derived the dominant contributions from dim-8 SMEFT operators to the meson mixing Hamiltonian, 
including one-loop QCD running results. In this section, we will apply these results into a LEFT master formula, 
which parameterizes NP contributions to the meson mixing parameters in terms of the eight LEFT WCs defined at $\Lambda_{\tt EW}$, 
and then calculate the bounds on the dim-8 SMEFT operators by comparing the SM predictions with experimental data.

\subsection{The LEFT master formula}

Neutral meson mixing is governed by the off-diagonal entries of the two-state Hamiltonian $\hat{H}=\hat{M}-i\hat{\Gamma}/2$, where the Hermitian $2\times2$ matrices $\hat{M}$ and $\hat{\Gamma}$ describe the off-shell and on-shell transitions respectively. The effective Hamiltonian ${\cal H}_{\Delta  F=2}^{ij}$ receives contributions from both the SM (${\cal H}_{\Delta  F=2}^{{\tt SM}, ij}$) and NP (${\cal H}_{\Delta  F=2}^{{\tt NP}, ij}$) effects.
Here, we focus on the off-shell contributions from new heavy particles, whose effects are parameterized by the higher-order effective interactions. Thus, the quantity $M_{12}$ that is related to the off-diagonal matrix elements of the effective Hamiltonian between the two meson states can be decomposed into the SM and NP parts,
\begin{align}
    M_{12}^{ij} ={\langle M^0|{\cal H}_{\Delta F=2}^{ij}|\overline{M^0}\rangle \over {2 M_{M^0}}} = [M_{12}^{ij}]_{\tt SM} + [M_{12}^{ij}]_{\tt NP},
    \label{eq:mm}
\end{align}
where $ij=ds,\,cu,\,db,\,sb$ correspond to the mixing of $K^0-\bar K^0$, $D^0-\bar D^0$, $B_d-\bar B_d$, and $B_s-\bar B_s$, respectively. These off-diagonal matrix elements are directly related to the experimentally measured quantities. 
For the four neutral meson systems, $M^0=\{K^0, D^0, B_s, B_d\}$, the measured mass differences $\Delta M_K$, $\Delta M_{B_i}$ ($i=d,\,s$), and $\Delta M_{D}$ are represented by 
\begin{align}
    \Delta M_{K^0} =2 \Re(M_{12}^{ds}),\quad \Delta M_{B_i} =2 |M_{12}^{ib}|,\quad \Delta M_{D^0} =2 |M_{12}^{cu}|.
\end{align} 
The SM prediction for $\Delta M_{K^0}$ reported by the recent lattice QCD calculation \cite{Wang:2022lfq} is, $(\Delta M_{K^0})_{\tt SM}=5.8(6)(2.3)\times 10^{-15}$ GeV, which has achieved a relatively small statistical error.
Meanwhile, its experimental value is $(\Delta M_{K^0})_{\tt exp}=3.484(6)\times 10^{-15}$ GeV \cite{ParticleDataGroup:2024cfk}. This result suggests consistency between the experimental measurement and the SM prediction, considering the finite lattice spacing errors in the calculation. For the $B$ meson mixing, the combined results from different measurements lead to 
$\Delta (M_{B_d})_{\tt exp}=3.336(12)\times 10^{-13}\,{\rm GeV}$ and $(\Delta M_{B_s})_{\tt exp}=1.1693(4)\times 10^{-11}\,{\rm GeV}$ \cite{HFLAV:2022esi,ParticleDataGroup:2024cfk}.
These results are consistent with the SM calculations within uncertainties. The latest SM predictions are $(\Delta M_{B_d})_{\tt SM}=3.521(138)\times10^{-13}$ GeV and $(\Delta M_{B_s})_{\tt SM}=1.1999(415)\times10^{-11}$ GeV \cite{Albrecht:2024oyn}, respectively, which have updated the previous calculations in \cite{Lenz:2019lvd,DiLuzio:2019jyq} with new CKM and hadronic matrix elements inputs. 
For the $D^0-\bar{D}^0$ mixing, the commonly used observables are
\begin{align}
x_{12} = {\Delta M_{D^0} \over \Gamma_{D^0}}, \quad y={\Delta \Gamma_{D^0} \over 2\Gamma_{D^0} },
\end{align}
with $\Gamma_{D^0}$ being the average total decay rate of the neutral $D$ mesons. The global fit of the HFLAV group gives $x=(0.407 \pm 0.044)\, \%$ which corresponds to $(\Delta M_{D^0})_{\tt exp}=6.56(76)\times 10^{-15}\,{\rm GeV}$ \cite{ParticleDataGroup:2024cfk}. The theoretical calculations for $D^0-\bar D^0$ mixing are not well established, and the predictions for $|x|$ and $|y|$ through different approaches  range from $\calO(10^{-5})$ to $\calO(10^{-2})$ without reliable error estimates \cite{Nierste:2009wg,Nelson:1999fg,Petrov:2003un,Chavez:2012xt,Lenz:2020awd}. Given the large uncertainties in SM calculations, there exist studies that attempt to attribute the experimental values of $D^0{\rm -}\bar{D}^0$ mixing as the NP effects \cite{Golowich:2007ka}. In \cref{tab:input}, we summarize the mass differences from the SM predictions and experimental measurements discussed above.  

\begin{table}
\center
\resizebox{\linewidth}{!}{
\renewcommand\arraystretch{1.7}
\begin{tabular}{|c|c|c|c|ccccc|c|}
\hline
\multirow{2}*{Meson}  & \multicolumn{2}{c|}{$\Delta M_{M^0}\,[\rm GeV]$} & \multicolumn{7}{c|}{SUSY basis}
\\
\cline{2-10}
& SM prediction & Experiment & $ij$ & $P_1^{ij}(\Lambda_{\tt EW})$ & $P_2^{ij}(\Lambda_{\tt EW})$ & $P_3^{ij}(\Lambda_{\tt EW})$ & $P_4^{ij}(\Lambda_{\tt EW})$ & $P_5^{ij}(\Lambda_{\tt EW})$ & units
\\
\hline
$K^0$ & $5.8(6)(2.3)\times 10^{-15}$ & $3.484(6)\times 10^{-15}$ & $ds$ & 0.102(2) & $-4.32(16)$ & $1.09(5)$ & $14.14(82)$ & $4.28(14)$ & $10^7\, \rm TeV^2$ 
\\
\hline
$D^0$ & $ 10^{-17}{\rm-}10^{-14} $ & $6.56(76)\times 10^{-15}$ & $cu$ & $0.54^{+0.17}_{-0.18}$ & $-2.11^{+0.65}_{-0.69}$ & $0.54^{+0.17}_{-0.18}$  &  $5.94^{+1.88}_{-1.96}$ &  $2.04^{+0.64}_{-0.67}$ & $10^7\, \rm TeV^2$  
\\
\hline
$B_d$ & $3.521(138)\times 10^{-13}$ & $3.336(12)\times 10^{-13}$ & $db$ & $2.67(10)$ & $-4.99(28)$ & $1.12(8)$ & $12.74(50)$ & $5.15(27)$ & $10^5\, \rm TeV^2$ 
\\
\hline
$B_s$ & $1.1999(415)\times 10^{-11}$ & $1.1693(4)\times 10^{-11}$ & $sb$ & $1.15(4)$ & $-2.24(13)$  & $0.51(3)$ & $5.22(21)$ & $2.10(9)$ & $10^4\, \rm TeV^2$ 
\\\hline
\end{tabular} }
\caption{Experimental values and SM predictions on $\Delta M_{M^0}$ and normalized hadronic matrix elements in the SUSY basis are shown. 
Due to large uncertainties in theoretical predictions on $\Delta M_{D^0}$, 
a range of values is extracted from the calculations in \cite{Nierste:2009wg,Nelson:1999fg,Petrov:2003un, Chavez:2012xt,Lenz:2020awd}.
}
\label{tab:input}
\end{table}

From now on, we focus on the NP term $[M_{12}^{ij}]_{\tt NP}$ in \cref{eq:mm}. Such NP contributions to the $\Delta F=2$ neutral meson mixing in the LEFT framework were numerically 
incorporated into a master formula \cite{Aebischer:2020dsw}, 
\begin{align}
\label{eq:for}
2[M_{12}^{ij}]_{\tt NP}^{\tt LEFT} = 
(\Delta M_{ij})_{\tt exp}  
\left[ \sum_{a=1}^5 P_a^{ij}(\Lambda_{\tt EW}) C_a^{ij}(\Lambda_{\tt EW})
+ \sum_{b=1}^3 P_b^{ij}(\Lambda_{\tt EW}) \tilde C_b^{ij}(\Lambda_{\tt EW})
\right],
\end{align} %
which is based on the effective Hamiltonian in \cref{eq:effH}. In this formula, the NP contribution is factorized into three independent pieces: the experimentally measured value $(\Delta M_{ij})_{\tt exp}$, the normalized hadronic matrix elements $P_a^{ij}(\Lambda_{\tt EW})$ defined at the electroweak scale $\Lambda_{\tt EW}$, and the NP contribution to the LEFT operators characterized by the WCs $C_a^{ij}(\Lambda_{\tt EW})$ and $\tilde C_b^{ij}(\Lambda_{\tt EW})$. The SMEFT matching results for the LEFT WCs are given in \cref{tab:matching} while the inputs for $(\Delta M_{ij})_{\tt exp}$ are collected in \cref{tab:input}.
The explicit definition for the term $P_a^{ij}(\Lambda_{\tt EW})$ is,
\begin{align}
    P_a^{ij}(\Lambda_{\tt EW})= {\langle M^0| \calO_a^{ij}(\Lambda_{\tt EW}) |\overline{M^0}\rangle  \over M_{M^0}(\Delta M_{ij})_{\tt exp}},
\end{align} 
where the hadronic matrix elements $\langle M^0| \calO_a^{ij}(\Lambda_{\tt EW})|\overline{M^0}\rangle $ are evaluated at $\Lambda_{\tt EW}=160$ GeV. As clarified in \cite{Aebischer:2020dsw}, $\langle M^0| \calO_a^{ij}(\Lambda_{\tt EW})|\overline{M^0}\rangle $ is obtained by taking into account the QCD running effects of the physical matrix element defined at a relevant hadronic scale $\mu_{\tt had}$, $\langle M^0| \calO_a^{ij}(\mu_{\tt had})|\overline{M^0}\rangle$ \cite{Bazavov:2017weg,Boyle:2017ssm,FLAG:2019iem,FermilabLattice:2016ipl,Dowdall:2019bea}. The scale dependence of $P_a^{ij}(\Lambda_{\tt EW})$ will be compensated by the same scale dependence of $C_a^{ij}(\Lambda_{\tt EW})$ and $\tilde C_b^{ij}(\Lambda_{\tt EW})$, ensuring that physical observables are scale-independent. 
The values of $P_a^{ij}(\Lambda_{\tt EW})$ for all of the $K^0$, $D^0$, and $B_{d(s)}$ meson mixing in the SUSY basis have been calculated in \cite{Aebischer:2020dsw}, and are also collected in \cref{tab:input}. In the table, we show the latest measurements for $\Delta M_{D^0}$ and $\Delta M_{B_{d,s}}$, which are slightly different from the old values used in that work. Accordingly, the numbers for $P_a^{ij}(\Lambda_{\tt EW})$ related to $D^0$ and $B_{d,s}$ get slightly changed.

\subsection{Numerical constraints}
\label{sec:numerical}

With the framework outlined above, it is straightforward to obtain by using \cref{eq:for} the ratio of the NP contribution to the mass difference over the experimental value. To constrain the dim-8 operators, we require the ratio to be less than $10\,\%$, i.e., $2[M_{12}^{ij}]_{\tt NP}^{\tt LEFT}/ (\Delta M_{ij})_{\tt exp}\lesssim 10\,\%$. This is a relatively conservative estimation, especially for the $B_{d,s}$ mesons, as both theoretical and experimental precisions are better than $10\,\%$ as shown in \cref{tab:input}. The experimental measurement for $K^0$ has achieved the $\calO(0.1\,\%)$ precision level, although the SM prediction still suffers from large systematic uncertainties. 
Since the SM predictions and measured values are consistent within errors, our analysis assumes that the observed neutral meson mixing is primarily due to the SM contributions. In the following, We proceed both with and without considering the RG effect to constrain the dim-$8$ SMEFT operators listed in \cref{tab:SMEFTd8d6}, and consider one operator a time. 

\begin{figure}[t]
\centering
\includegraphics[width=0.49\linewidth]{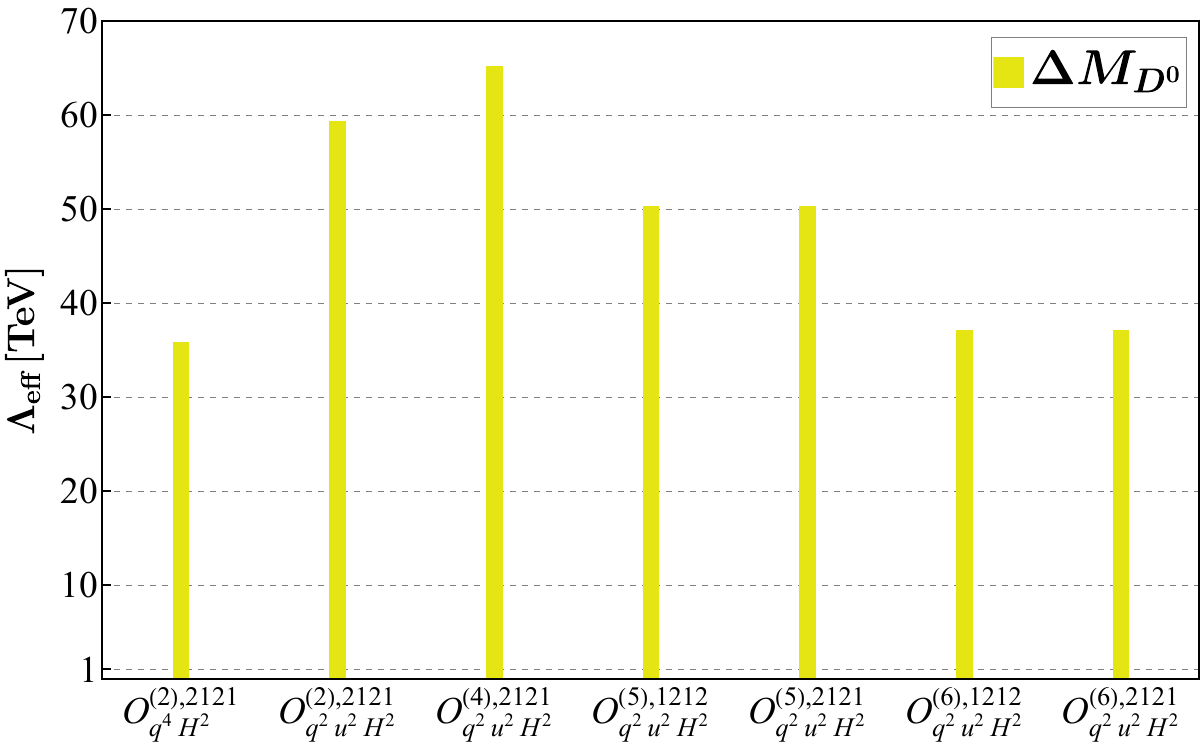}\,
\includegraphics[width=0.49\linewidth]{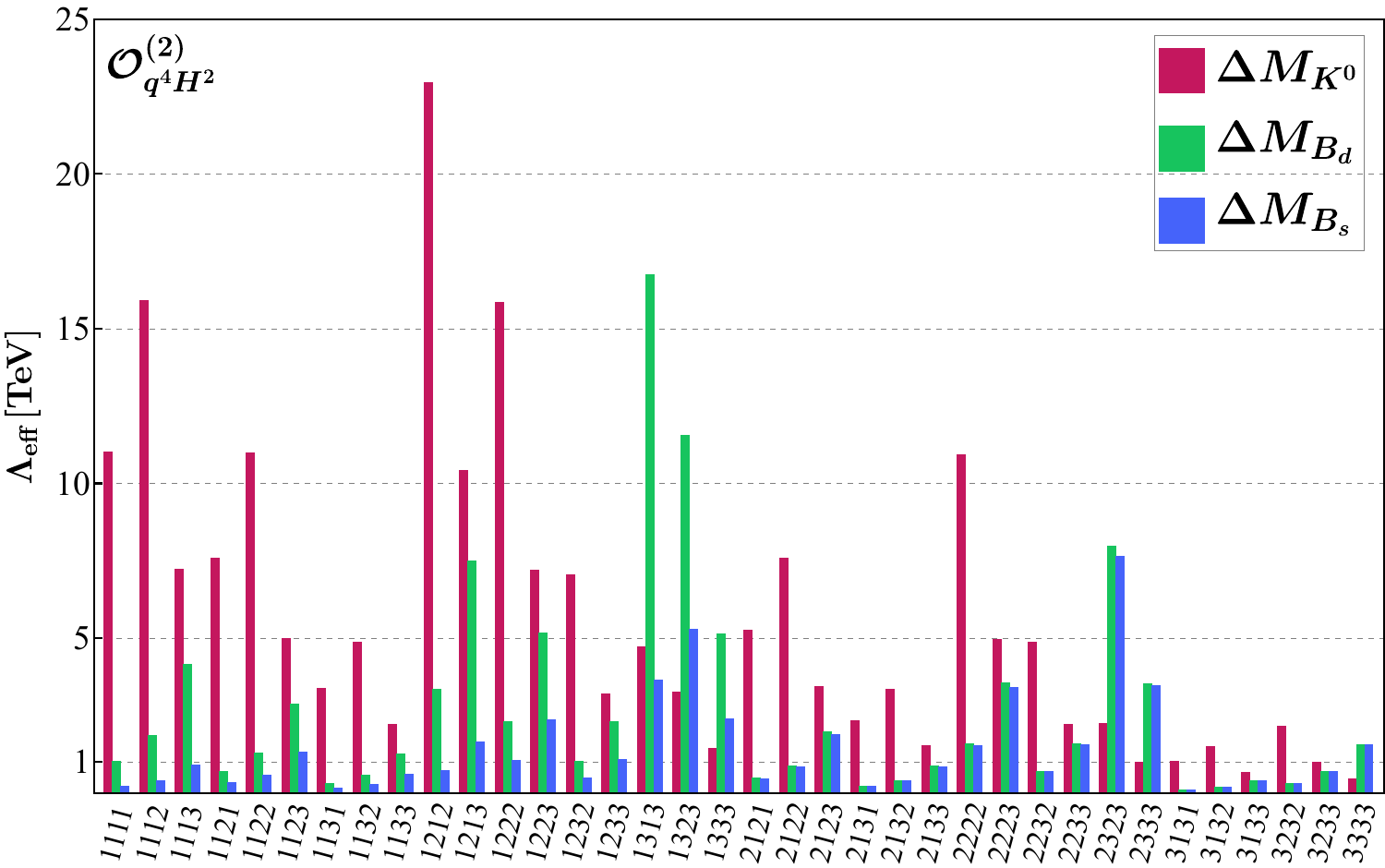}
\\
\vspace{5mm}
\includegraphics[width=0.49\linewidth]{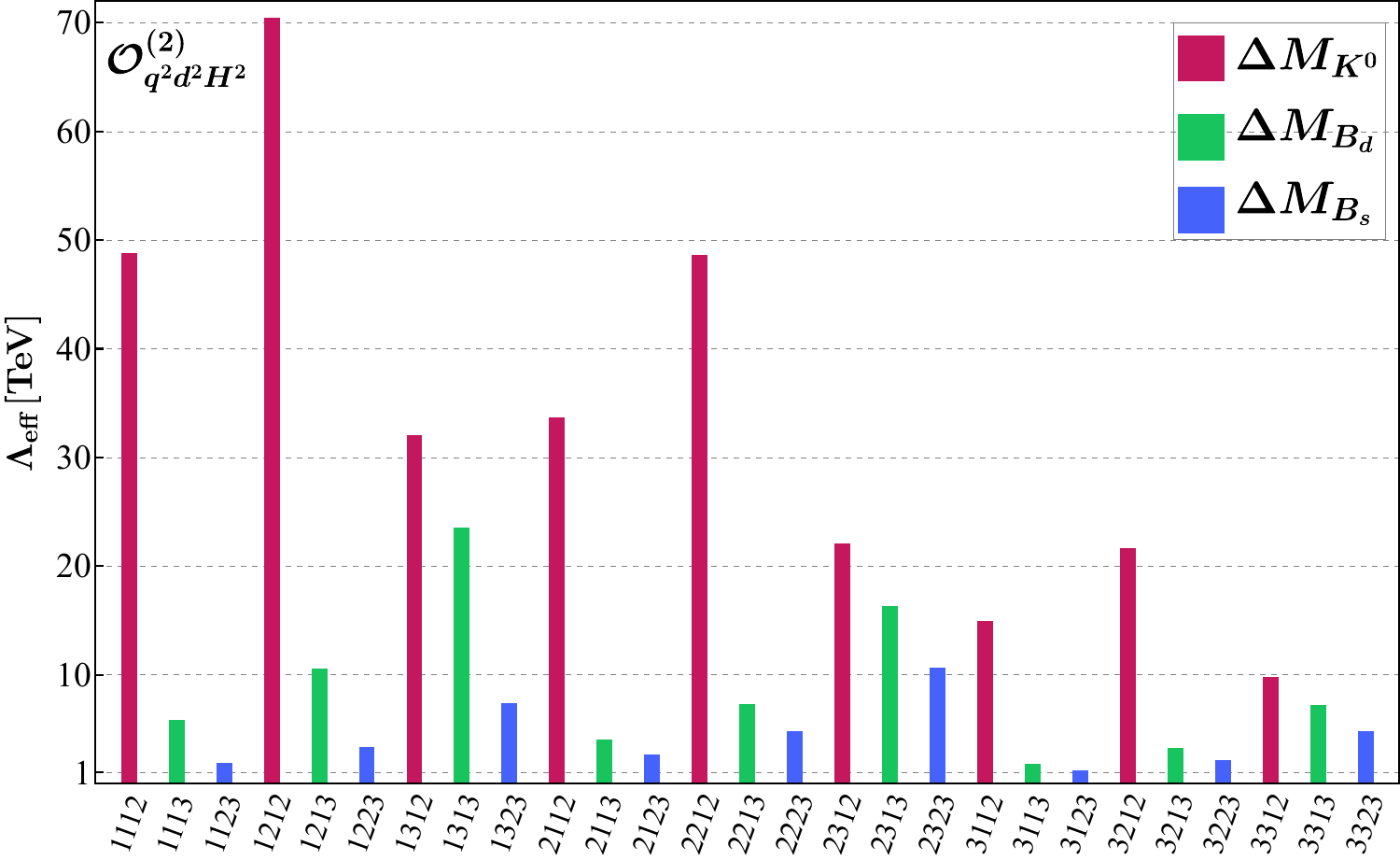}\,
\includegraphics[width=0.49\linewidth]{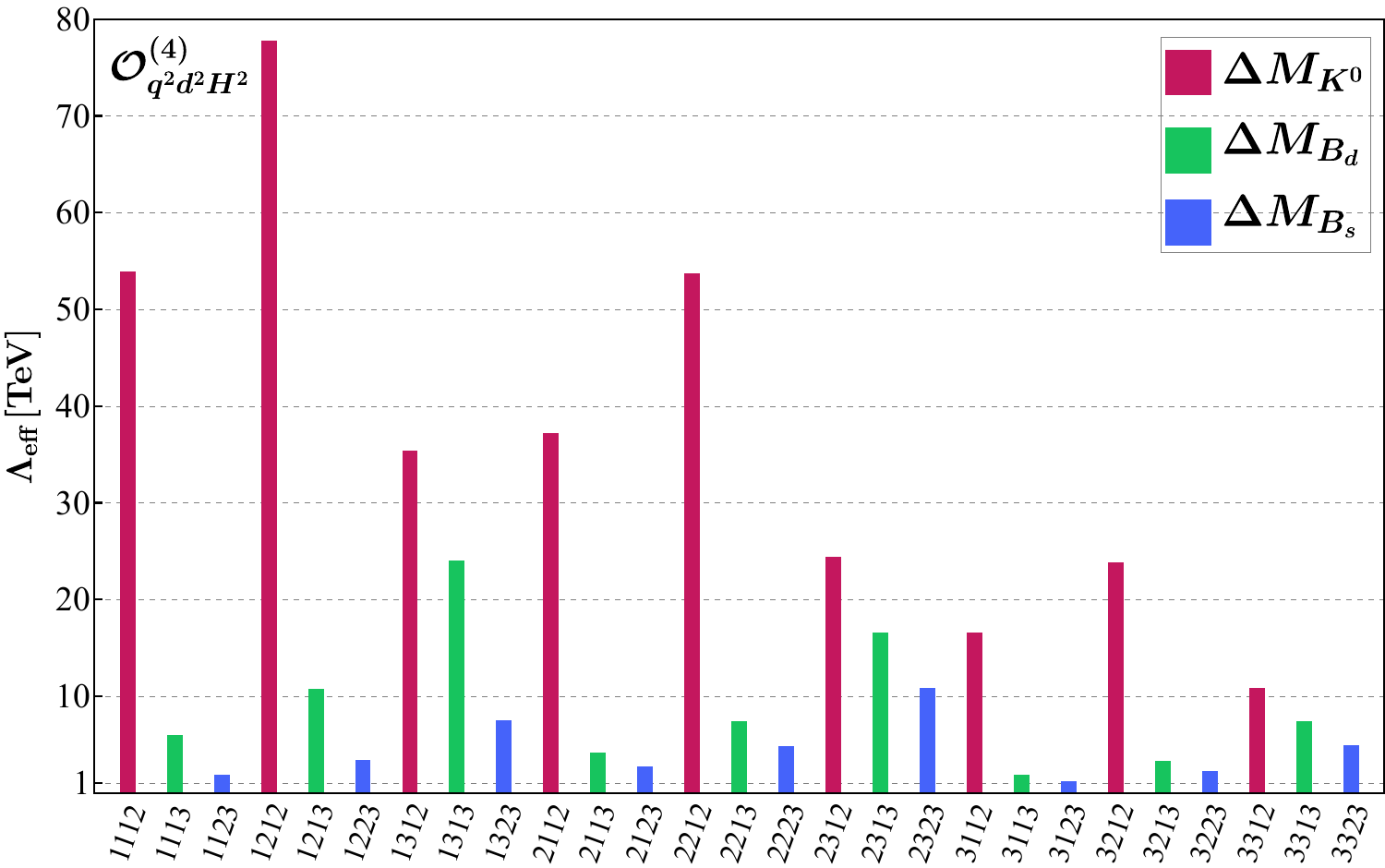}
\\
\vspace{5mm}
\includegraphics[width=0.49\linewidth]{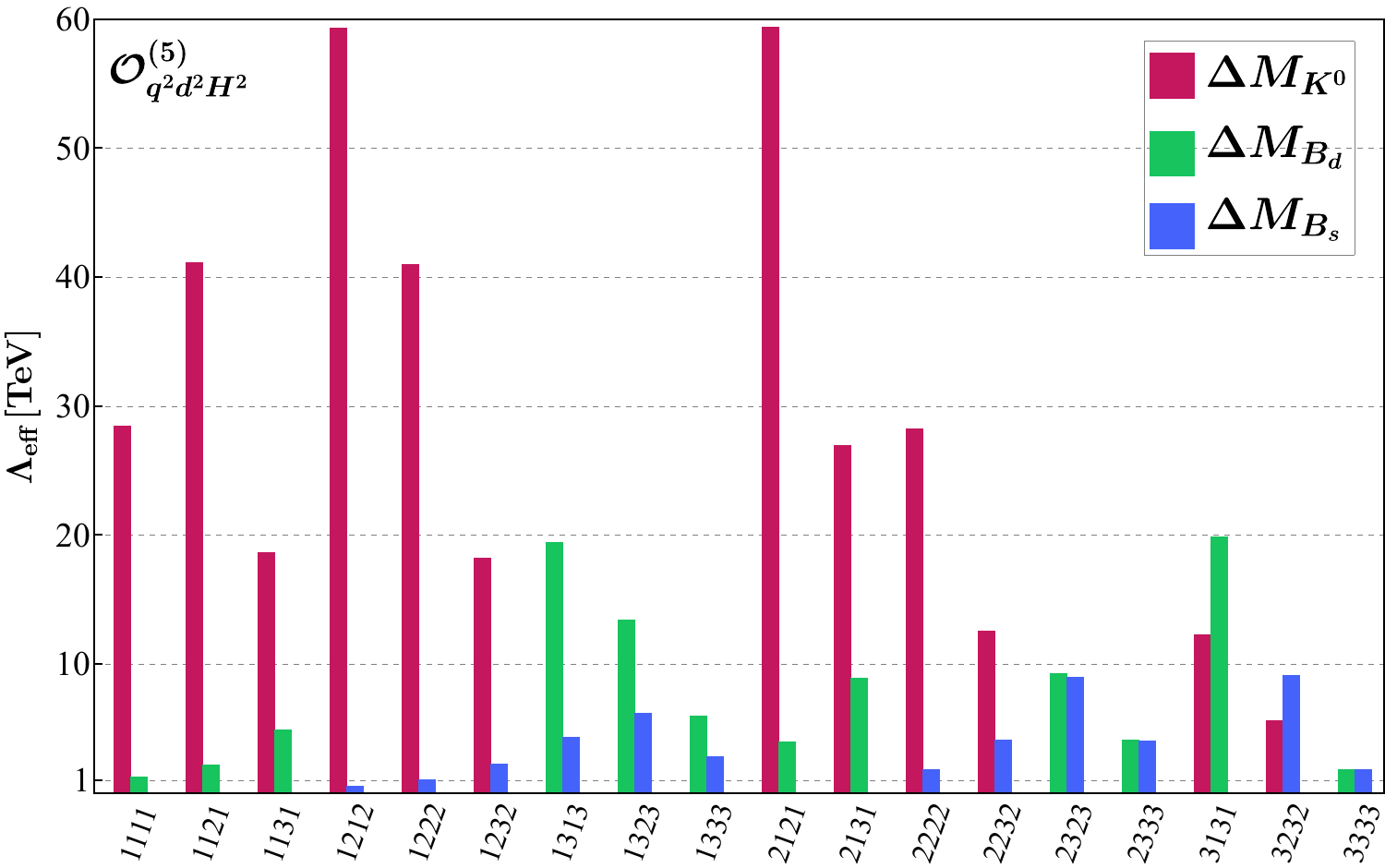}\,
\includegraphics[width=0.49\linewidth]{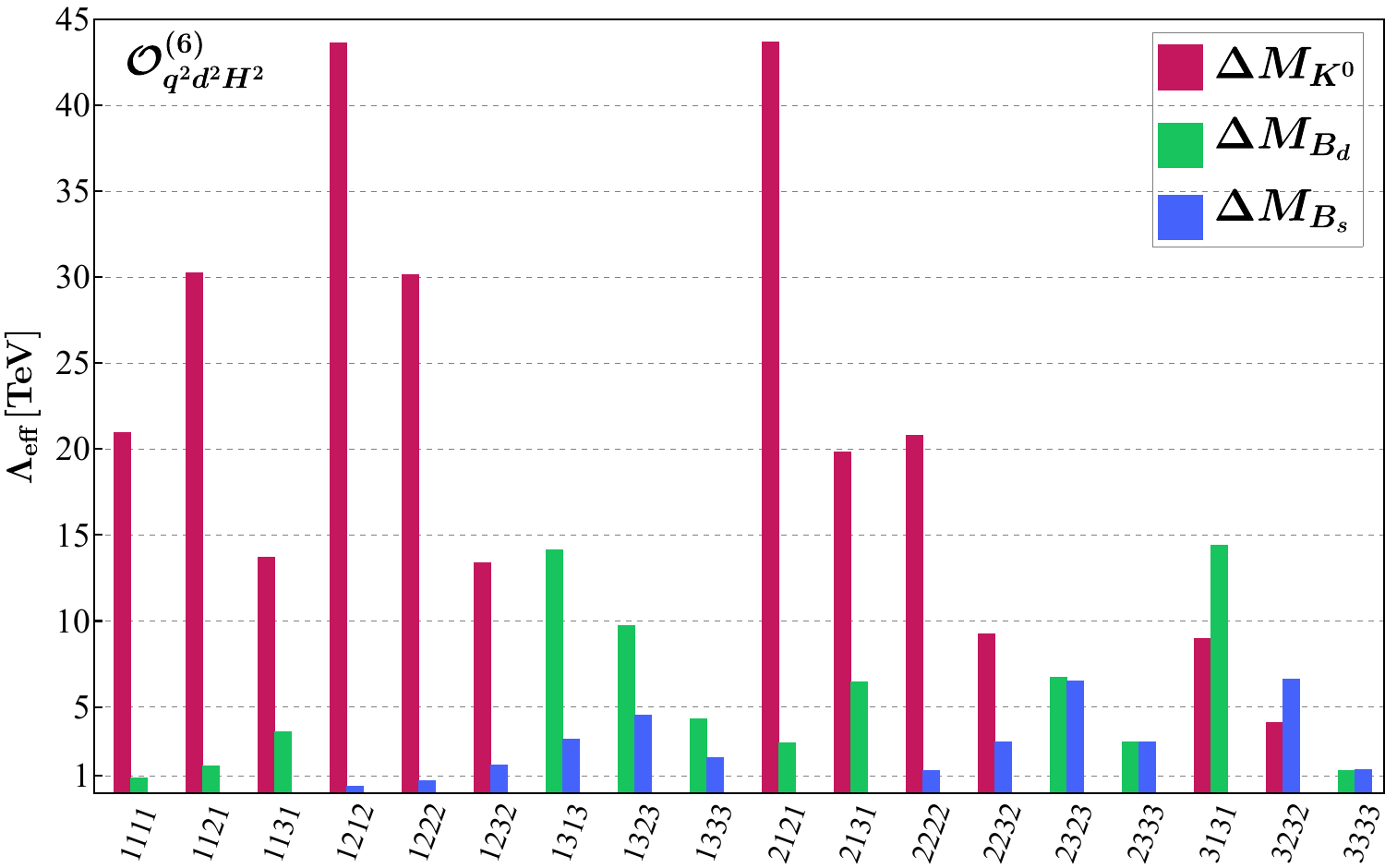}
\caption{The lower bound on the effective scale that can be probed by Type-II dim-8 operators. }
\label{fig:cons-noRGE}
\end{figure}

\noindent
{\bf The constraints at $\mu=\Lambda_{\tt EW}$: } 
We begin by setting the bounds on the WCs $C_{i}(\Lambda_{\tt EW})\equiv\Lambda_{\tt eff}^{-4}$ of dim-8 operators at the scale $\mu=\Lambda_{\tt EW}$ using the tree-level matching results provided in \cref{tab:matching}.
By investigating the $K^0$, $B_d$, $B_s$, and $D^0$ meson mixing parameters, various flavor structures can be explored. 
The results for the Type-II dim-8 operators are shown in \cref{fig:cons-noRGE}. 
Among the operators, ${\cal O}_{q^4 H^2}^{(5)}$ does not contribute due to its antisymmetry under the flavor interchange of its two currents. We found that the bounds from $\Delta M_{K^0}$ and $\Delta M_{D^0}$ are the most stringent, 
with the lower bounds exceeding $10$ TeV for most flavor structures. 
This can be explained by the relatively larger values for $P_a^{ds}$ and $P_a^{cu}$, 
which are enhanced by the factors $1/(M_{K^0} \Delta (M_{K^0})_{\tt exp})$ and $1/(M_{D^0} \Delta (M_{D^0})_{\tt exp})$, respectively. In the following, we first discuss the bounds on these Type-II  operators one by one.

{\bf Operator ${\cal O}_{q^4 H^2}^{(2)}$:}
The operator ${\cal O}_{q^4 H^2}^{(2)}$ contributes to both down-type $\Delta F=2$ mixing (where the corresponding observables are $\Delta M_{K^0}$ and $\Delta M_{B_{d,s}}$) and up-type $\Delta F=2$ mixing (specifically the observable $\Delta M_{D^0}$). Owing to the involvement of CKM matrix elements, all flavor combinations of ${\cal O}_{q^4 H^2}^{(2),xyzw}$ contribute to the down-type meson mixing. 
As can be seen from \cref{tab:matching} the contributions from ${\cal O}_{q^4 H^2}^{(2),xyzw}$ are the same with $x\leftrightarrow z$ or $y\leftrightarrow w$ or both, therefore we only display in \cref{fig:cons-noRGE} (top-right panel) the 36 independent constraints with $x\leq z$ and $y\leq w$. 
The most sensitive ones to $\Delta M_{K^0}$, $\Delta M_{B_d}$, and $\Delta M_{B_s}$ are ${\cal O}_{q^4 H^2}^{(2),1212}$, ${\cal O}_{q^4 H^2}^{(2),1313}$, and ${\cal O}_{q^4 H^2}^{(2),2323}$, respectively, as they receive the largest diagonal elements in the CKM matrix for each corresponding channel. Since we work with the so-called up-quark flavor basis, only the operator ${\cal O}_{q^4 H^2}^{(2),2121}$ leads to $D^0-\bar{D}^0$ mixing, and the lower limit on its effective scale can reach $\sim 35$ TeV.

{\bf Operator ${\cal O}_{q^2 d^2 H^2}^{(2,4)}$:} We found that the behavior of the operators ${\cal O}_{q^2 d^2 H^2}^{(2,4),xyzw}$ is very similar. The flavor structures $(z,w)=(1,2),\,(1,3),\,(2,3)$ are respectively sensitive to $\Delta M_{K^0}$, $\Delta M_{B_d}$, and $\Delta M_{B_s}$. Among all Type-II dim-8 operators the ${\cal O}_{q^2 d^2 H^2}^{(2,4)}$ give the largest contributions to $\Delta M_{K^0}$ and the translated lower bound on $\Lambda_{\rm eff}$ can reach $\sim 80$ TeV,
with the enhancements coming from $P_{4,5}^{ds}(\Lambda_{\tt EW})$. 

{\bf Operator ${\cal O}_{q^2 d^2 H^2}^{(5,6)}$:} Considering the observables $\Delta M_{K^0}$, $\Delta M_{B_d}$, and $\Delta M_{B_s}$, 
the operators ${\cal O}_{q^2 d^2 H^2}^{(5,6),xyzw}$ with flavor structures $y=w$ can be constrained. 
Furthermore, since the operators are symmetric under $(x,y) \leftrightarrow (z,w)$, 
there are $18$ independent flavor assembles with $x\leq z$ and $y = w$ as shown in \cref{fig:cons-noRGE}. 

The observable $\Delta M_{D^0}$ is specifically sensitive to the up-type operators, 
$\calO_{q^2 u^2 H^2}^{(2,4,5,6)}$ with flavor structures $(x,y,z,w)=(1,2,1,2)$ or $(x,y,z,w)=(2,1,2,1)$. 
The constraints on them together with the one on $\calO_{q^4 H^4}^{(2),2121}$ are shown in the  top-left panel of \cref{fig:cons-noRGE}. 
The resulting lower bounds on $\Lambda_{\rm eff}$ vary from $\sim 30$ TeV to $\sim 65$ TeV. 

Now we turn to the Type-I operators. Without considering RG effects, 
the contributions from most of the Type-I dim-8 operators are either the same as or opposite in sign to those from the Type-II operators. 
Since the operators are treated one by one, the results for these Type-I operators are identical to the corresponding Type-II operators, 
and we list the relation of an equal constraint by the notation $\sim$: 
$\calO^{(1,3),xyzw}_{q^4 H^2}\sim\calO^{(2),xyzw}_{q^4 H^2}$, 
$\calO^{(3),xyzw}_{q^2 \psi^2 H^2}\sim\calO^{(4),xyzw}_{q^2 \psi^2 H^2}$, 
$\calO^{(1),xyzw}_{q^2 \psi^2 H^2}\sim\calO^{(2),xyzw}_{q^2 \psi^2 H^2}$,
where $\psi=u,d$ represent both the up-type and down-type quark fields. 
The only exceptions are operators involving four right-handed quarks, 
i.e., $\calO_{d^4 H^2}$ and $\calO_{u^4 H^2}$, and we list the bounds in the third column of \cref{tab:cons:ind}.

\begin{table}
\center
\resizebox{0.8 \linewidth}{!}{
\renewcommand\arraystretch{1.1}
\begin{tabular}{|c|c|c|c|}
\hline
Operator & Observable & Bounds on $\Lambda_{\rm eff}$ [TeV] & Bounds on $c_i$ at $\Lambda_{\tt NP}=5\,\rm TeV$
\\
\hline
\hline
$\calO_{d^4 H^2}^{1212}$ & $\Delta M_{K^0}$ & $23.57$ & $2.25\times 10^{-3}$
\\
$\calO_{d^4 H^2}^{1313}$ & $\Delta M_{B_d}$ & $16.86$ & $8.60\times 10^{-3}$ 
\\
$\calO_{d^4 H^2}^{2323}$ & $\Delta M_{B_s}$ & $7.68$  &  $0.20$
\\
$\calO_{u^4 H^2}^{2121}$ & $\Delta M_{D^0}$ & $35.75$  &  $4.25\times 10^{-4}$ 
\\
\hline
\end{tabular} 
}
\caption{ Lower (upper) bounds on the effective scale $\Lambda_{\rm eff}$ 
(Wilson coefficient $c_i$ at $\Lambda_{\tt NP}=5\,\rm TeV$) of the Type-I dim-8 operators $\calO_{d^4 H^2}$
and $\calO_{u^4 H^2}$ without (with) including RG effects in the third (fourth) column.}
\label{tab:cons:ind}
\end{table}

\begin{figure}
\centering
\includegraphics[width=0.49\linewidth]{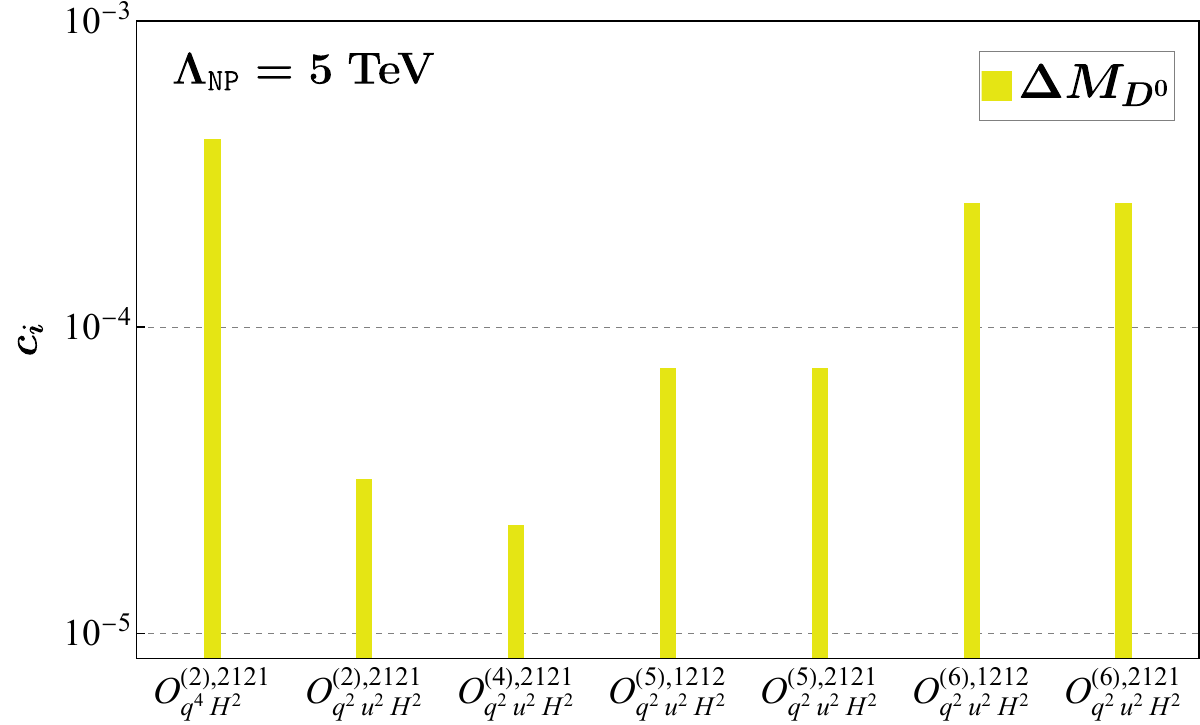}
\,
\includegraphics[width=0.49\linewidth]{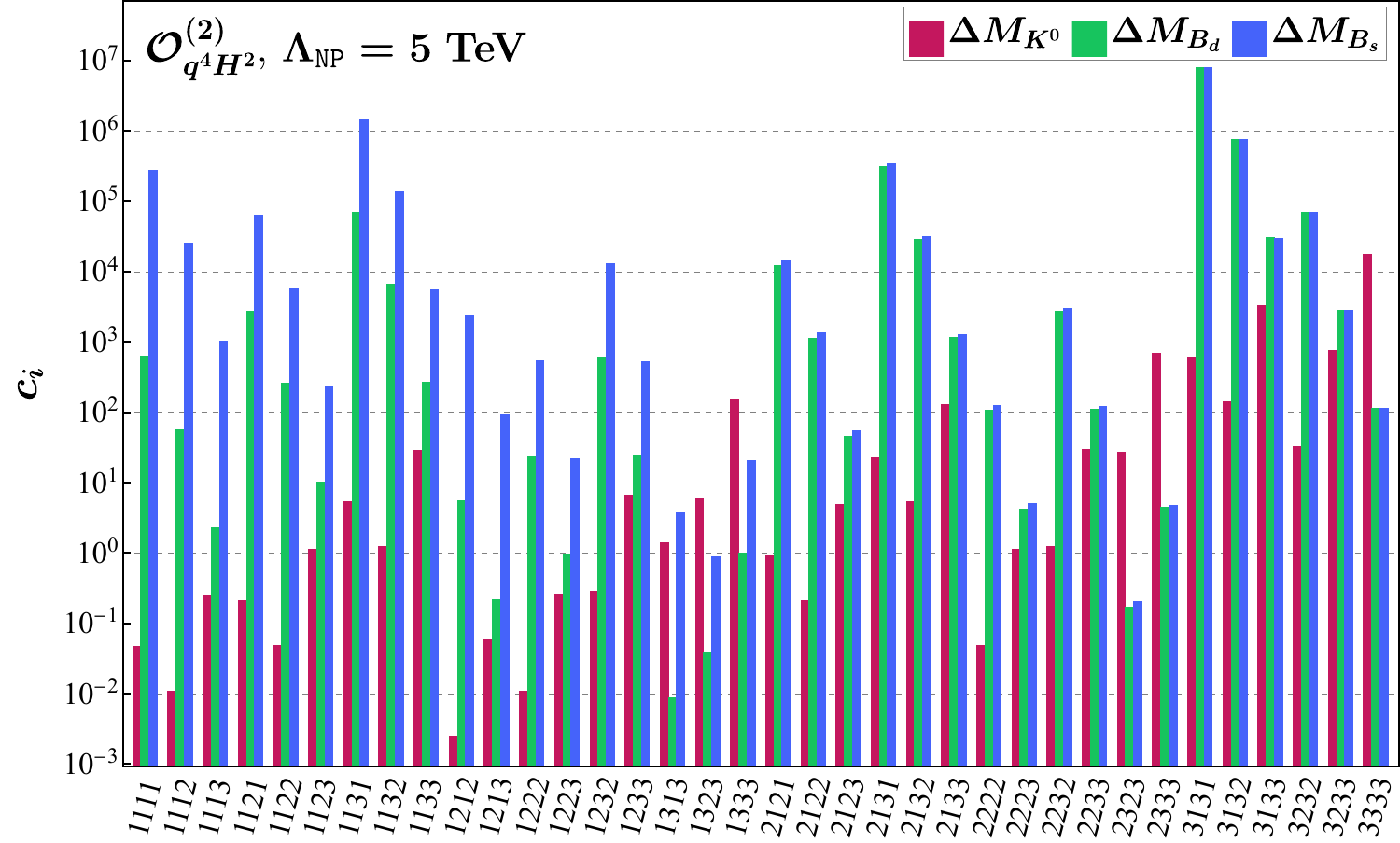}
\\
\vspace{5mm}
\includegraphics[width=0.49\linewidth]{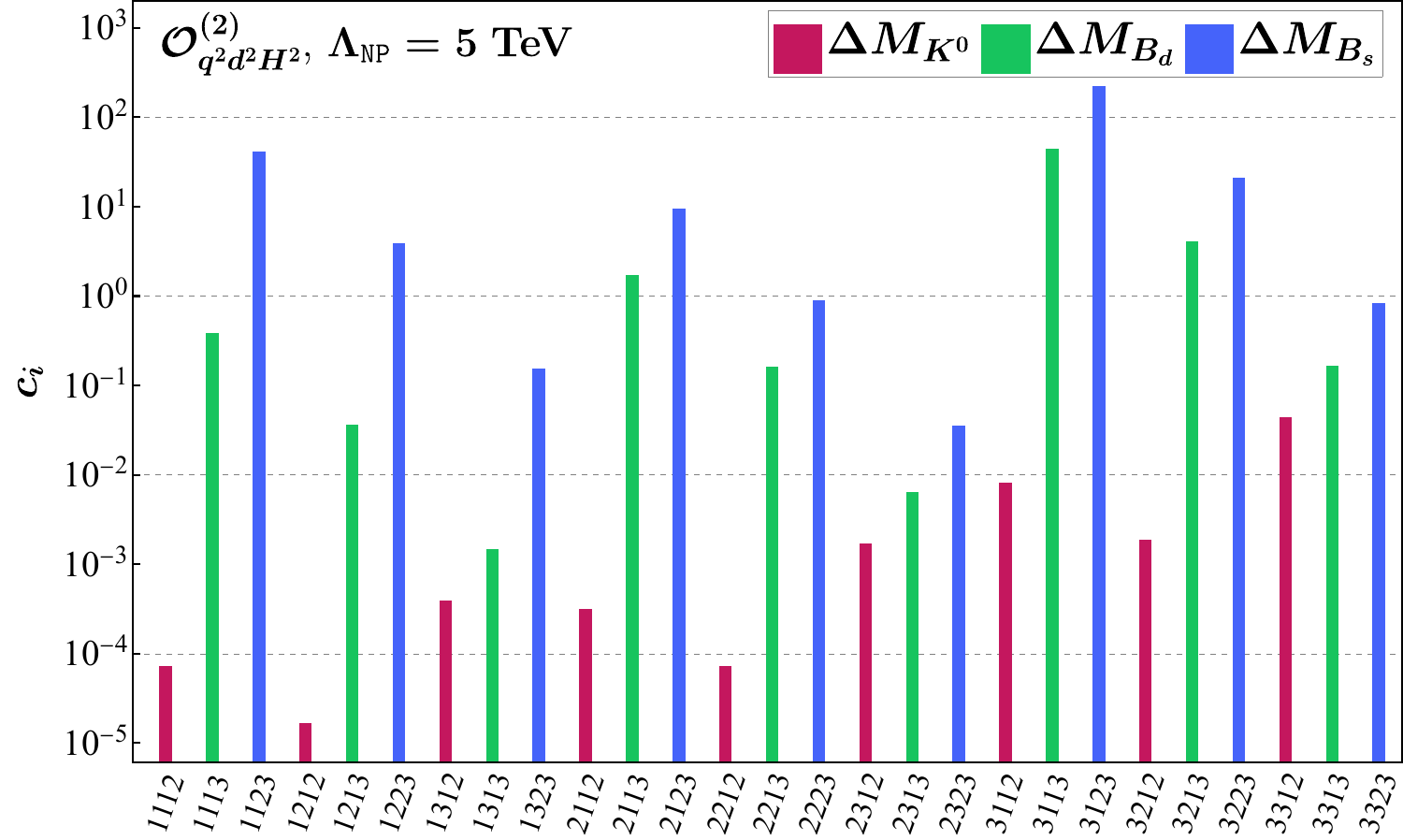}
\,
\includegraphics[width=0.49\linewidth]{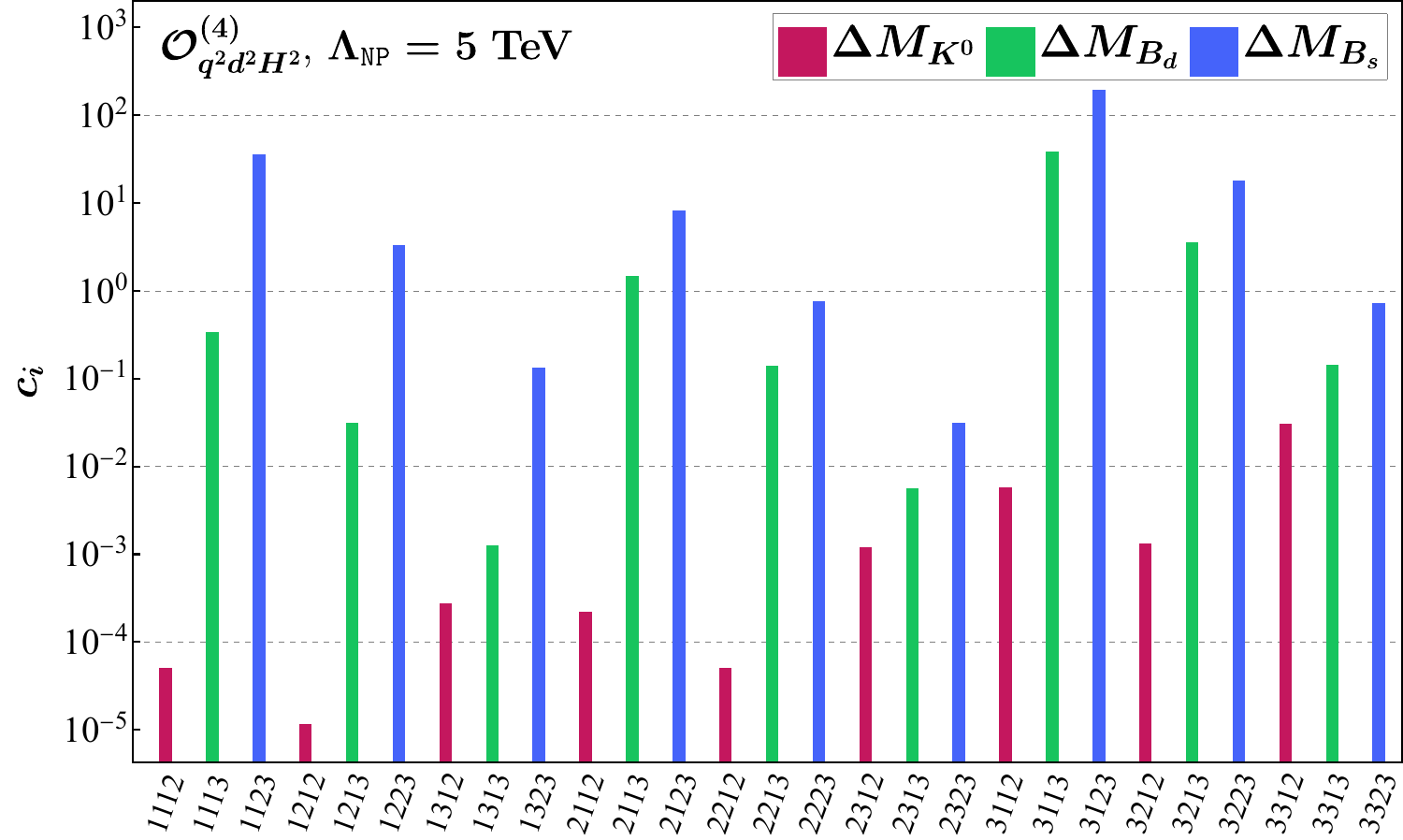}
\\
\vspace{5mm}
\includegraphics[width=0.49\linewidth]{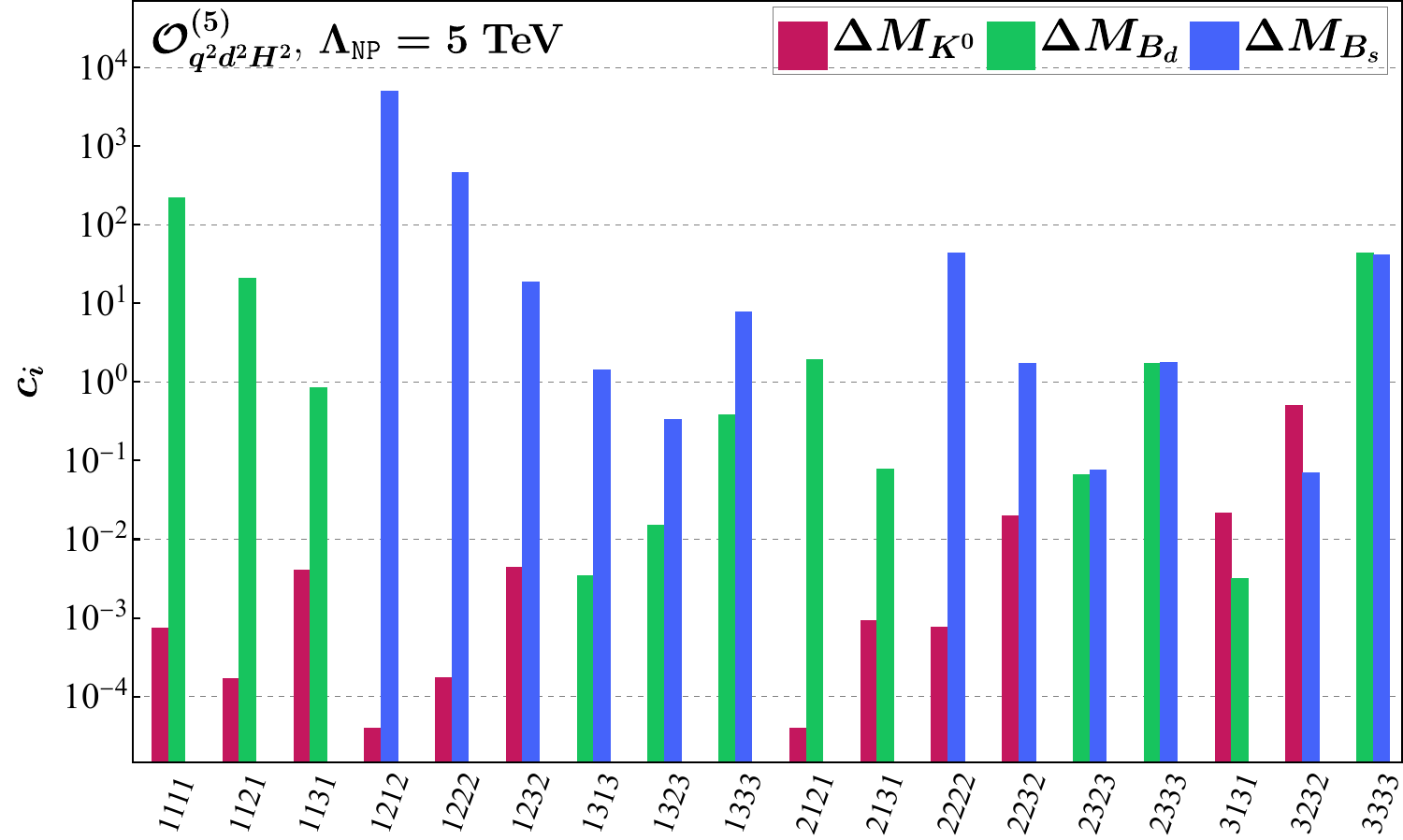}
\,
\includegraphics[width=0.49\linewidth]{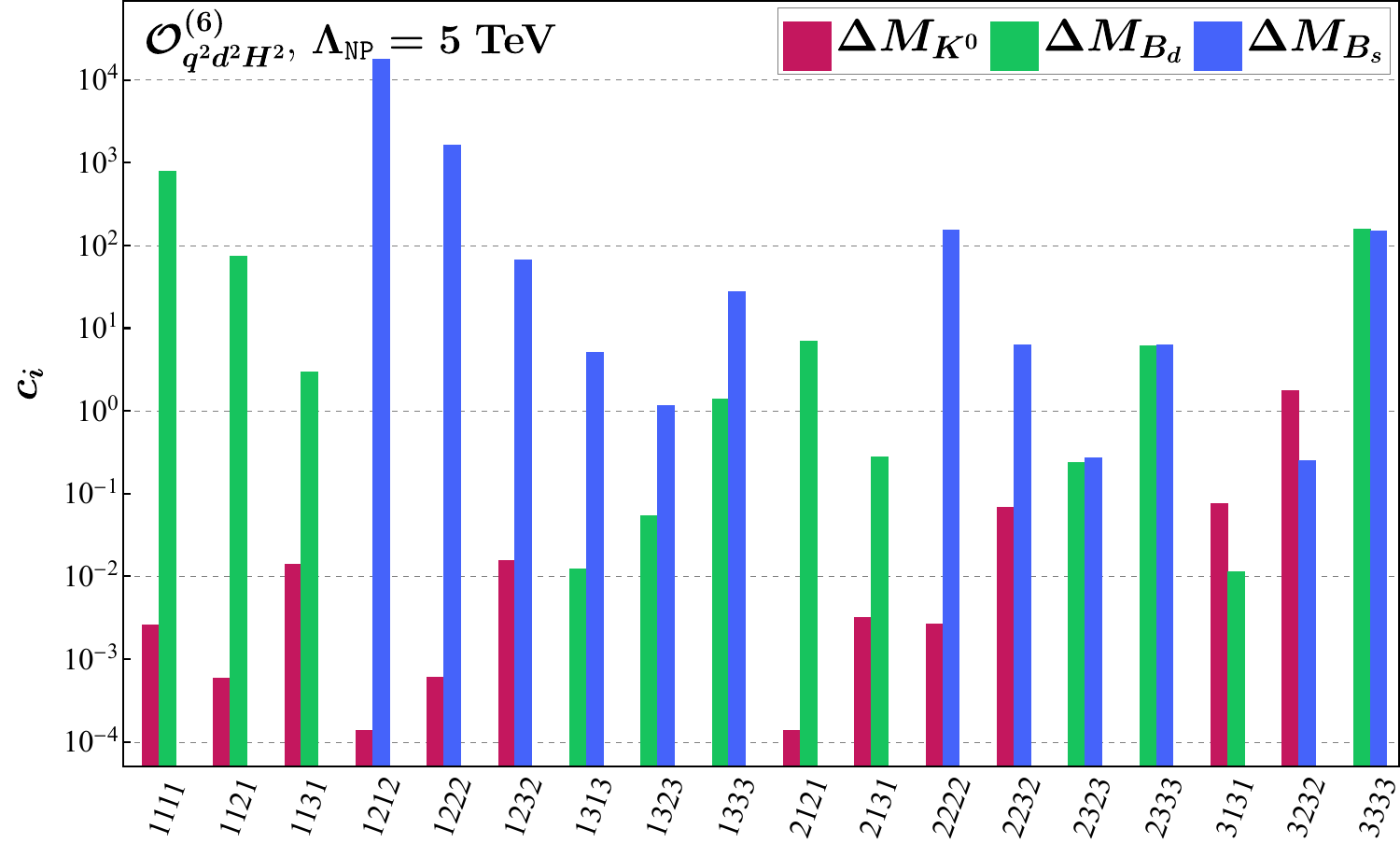}
\caption{ The upper bounds on dimensionless couplings at the scale $\mu=\Lambda_{\tt NP}=5~\textrm{TeV}$
for the Type-II dim-8 operators with RG effects incorporated. }
\label{fig:cc-RGE}
\end{figure}

%
\noindent
{\bf The constraints at $\mu=\Lambda_{\tt NP}$:~}  
We now consider the constraints on the WCs defined at a NP scale $\Lambda_{\tt NP}=5\,\rm TeV$ which we choose to be the same as that in \cite{Aebischer:2020dsw}. In doing so, we define a dimensionless coupling for each WC via $C_i \equiv c_i /\Lambda_{\tt NP}^4$ and then set a bound on $c_i$ upon incorporating the RG effects from $\mu=\Lambda_{\tt EW}$ to $\mu=\Lambda_{\tt NP}$. The resulting upper bounds on $c_i(\Lambda_{\tt NP})$ for Type-II operators are shown in \cref{fig:cc-RGE}. We find that the QCD RG effects maintain the relationships between the corresponding Type-I and Type-II operators, as implied in \cref{tab:matching}. This property leads to the same bounds on $c_i\,(\Lambda_{\tt NP})$ for the Type-I operators $\calO_{q^4 H^2}^{(1,3)}$ and $\calO_{q^2\psi^2 H^2}^{(1,3)}$ with $\psi=u,d$ as for the corresponding Type-II operators. The bounds on $\calO_{d^4 H^2}$ and $\calO_{u^4 H^2}$ are displayed in the fourth column of \cref{tab:cons:ind}. 
After the RG effects, the contributions from some operators are enhanced by a factor ranging from $1.06$ to $1.54$ (e.g., $\calO_{q^2 \psi^2 H^2}^{(1,2,3,4,5,6)}$), while the contributions from others are suppressed by a factor about $0.9$ (e.g., $\calO_{q^4  H^2}^{(1,2,3)}$ and $\calO_{\psi^4  H^2}$).

Compared to the studies aimed at exploring dim-8 SMEFT operators at colliders, e.g., \cite{Dawson:2024ozw,Boughezal:2022nof,Corbett:2023qtg}, our findings indicate that the analysis of $\Delta F=2$ neutral meson mixing offers the potential for significantly greater sensitivity. This suggests that NP could be more effectively uncovered or excluded with improvements in accuracy, particularly in SM calculations, within the neutral meson systems.

\section{An example of a UV model}
\label{sec:UVmodel}

An interesting model that can generate the relevant dim-8 but not dim-6 operators at leading order is provided in \cite{He:2024iju}, which was built to explain the potential anomaly in $B^+\to K^+\nu\bar\nu$ from the recent Belle II result \cite{Belle-II:2023esi}. The model introduces a light scalar field $\phi\sim(\mathbf{1},\mathbf{1},0)$ as a dark matter (DM) candidate, two heavy vector-like quarks $Q \sim(\mathbf{3},\mathbf{2},1/6)$ and $D \sim(\mathbf{3},\mathbf{1},-1/3)$ with masses $m_Q$ and $m_D$, respectively. The three numbers in parentheses indicate their representations and charges in the SM gauge group $\rm SU(3)_{\rm c}\times SU(2)_{\rm L}\times U(1)_{\rm Y}$. All these new fields ($\phi$, $Q_{L,R}$ and $D_{L,R}$) are odd under an imposed $\mathbb{Z}_2$ symmetry that stabilizes the DM particle $\phi$. Because of the $\mathbb{Z}_2$ symmetry, the vector-like quarks do not mix with the SM quarks, but they mix among themselves after electroweak symmetry breaking. The relevant new Yukawa interactions are 
\begin{align}
\mathcal{L}_{\tt Yukawa}^{\tt NP} & = 
y_{q}^x \bar{q}_{x} Q_{R} \phi + y_{d}^x \bar{D}_L d_{x} \phi 
- y_1 \bar{Q}_L D_R H - y_2 \bar{Q}_R D_L H + {\rm h.c.}\;, 
\end{align}
where $y_q^x, y_d^x, y_{1,2}$ are new Yukawa couplings with the superscript $x$ denoting the generation of quarks. The rephasing freedom of the two heavy quark fields can be used to choose $y_1$ and one of the Yukawa couplings $y_q^x$ to be real without loss of generality. All other Yukawa couplings generally are complex. 

Now we consider the contribution to the meson mixing in this model.
Due to the $\mathbb{Z}_2$ symmetry, there are no tree-level matching contributions to 
the SMEFT operators of any dimension. 
The non-trivial contributions to the SMEFT interactions first appear at one-loop level. But as pointed out in \cite{He:2024iju}, due to the chirality mismatch of the quark pairs from the one-loop diagrams with four-quark legs, the dim-6 SMEFT interactions related to meson mixing are absent in the model. The leading-order non-trivial contributions come from the dim-8 operators consisting of four quarks and two Higgs fields that are generated at one-loop order. By calculating the relevant one-loop diagrams with 
four-quark-two-Higgs external fields, we have the following one-loop matching results, 
\begin{align}
\label{eq:1Lmatching}
{\cal L}_{q^4 H^2}^{\tt SMEFT,(1)}= [ C_{q^2d^2H^2}^{(5),xyzw} (\bar q_x d_y H)(\bar q_z d_w H) +{\rm h.c.} ]
+ C_{qdqdH^2}^{xyzw} (\bar q_x d_y H)(H^\dagger \bar d_z q_w),
\end{align}
where the WCs are
\begin{align}
C_{q^2d^2H^2}^{(5),xyzw}  = - { y_q^x y_d^y y_q^z y_d^w y_1^2 \over 16\pi^2 m_Q^2 m_D^2  }( L_Q+  h_1), \quad 
C_{qdqdH^2}^{xyzw}  = - { y_q^x y_d^y  y_d^{z*} y_q^{w*}  |y_1|^2 \over 8\pi^2 m_Q^2 m_D^2}( L_Q+  h_2).
\end{align}
Here the loop functions are
\begin{subequations}
\begin{align}
h_1 & = 
{1\over (1-x)^3}\left\{ 2-4x + 4 x^2 - 2x^3 - (1- 3x)\ln x
\right.
\nonumber
\\
&\left. + 2 \sqrt{x} (1 - x^2 + 2 x\ln x)z 
+  x [2- 2x + (1+x) \ln x]z^2 \right\}, 
\\
h_2 & = h_1|_{z\to \Re[z], z^2 \to |z|^2}, 
\end{align}
\end{subequations}
with $x \equiv m_D^2/ m_Q^2$ and $z\equiv y_2/y_1$.
In the degenerate heavy mass limit of $x\to 1$, $h_1 \to (17+ 4 z - z^2)/6$. $L_Q = {1/ \epsilon} + \ln(4\pi\mu^2/ m_Q^2)- \gamma_{\rm E}$ is a divergent piece, which is canceled by the one-loop contribution involving double insertions of the tree-level matching result in the SMEFT extended with the DM field $\phi$,
\begin{align}
\label{eq:phiSMEFT}
\mathcal{L}_{\phi^2 q^2}^{\tt \phi SMEFT,(0)} & = 
C_{qdH\phi^2}^{xy} \mathcal{O}_{qdH\phi^2}^{xy}
+C_{quH\phi^2}^{xy} \mathcal{O}_{quH\phi^2}^{xy}
+\mathrm{h.c.}\;,
\end{align}
where the effective operators $\mathcal{O}_{qdH\phi^2}^{xy} = (\bar{q}_x d_y H)\, \phi^2$ and 
$\mathcal{O}_{quH\phi^2}^{xy}  = (\bar q_{x} u_{y} \tilde H) \phi^2$ with $\tilde H = \epsilon H^*$. The matching results for the WCs $C_{qdH\phi^2}$ and $C_{quH\phi^2}$ can be found in \cite{He:2024iju}. Thus, we will drop $L_Q$ in the following numerical analysis.

The second operator in \cref{eq:1Lmatching} can be converted into the basis operators
given in \cref{tab:SMEFTd8d6} by Fierz and group identities, 
\begin{align}
- 2 (\bar q_x d_y H)(H^\dagger \bar d_z q_w) 
 = {1\over 2 N_c }\calO_{q^2 d^2 H^2}^{(1),xwzy}+ {1\over 2 N_c }\calO_{q^2 d^2 H^2}^{(2),xwzy} 
+ \calO_{q^2 d^2 H^2}^{(3),xwzy} + \calO_{q^2 d^2 H^2}^{(4),xwzy}.
\end{align}
Due to the absence of tree-level matching here, our matching results are not affected by potential shifts via the one-loop insertion of the tree-level contributions of the evanescent operators \cite{Aebischer:2022aze,Fuentes-Martin:2022vvu}.
Thus, in the basis given in \cref{tab:SMEFTd8d6}, five relevant dim-8 operators are generated with their WCs being
\begin{align}
& C_{q^2d^2H^2}^{(5),xyzw}  = - { y_q^x y_d^y y_q^z y_d^w y_1^2 \over 16\pi^2 m_Q^2 m_D^2  }h_1,   \quad 
C_{q^2d^2H^2}^{(3),xyzw}  = { y_q^x y_q^{y*} y_d^{z*} y_d^w  |y_1|^2 \over 16\pi^2 m_Q^2 m_D^2}h_2,
\nonumber
\\
& C_{q^2d^2H^2}^{(1),xyzw} =\frac{1}{6} C_{q^2d^2H^2}^{(3),xyzw},\, 
C_{q^2d^2H^2}^{(2),xyzw} =\frac{1}{6} C_{q^2d^2H^2}^{(3),xyzw},\,
C_{q^2d^2H^2}^{(4),xyzw} = C_{q^2d^2H^2}^{(3),xyzw}.
\end{align}
If we choose the same NP scale as in \cite{He:2024iju} $m_Q=m_D=3\,\rm TeV$ that can accommodate the recent $B\to K\nu\bar\nu $ anomaly \cite{Belle-II:2023esi}, the nonvanishing LEFT WCs from the matching at $\Lambda_{\tt EW}$ become,
\begin{align}
& C_{2,dd}^{ij}(\Lambda_{\tt EW})  =
 1.14 \times {v^2\over 2} C_{q^2 d^2 H^2}^{(5),jiji}(3\,\rm TeV), 
\nonumber
\\
& C_{3,dd}^{ij}(\Lambda_{\tt EW})  = -  0.09 \times {v^2\over 4} C_{q^2 d^2 H^2}^{(5),jiji}(3\,\rm TeV),
\nonumber
\\
& C_{4,dd}^{ij}(\Lambda_{\tt EW})  = - 1.58 \times  {v^2 \over2}  C_{q^2 d^2 H^2}^{(3),ijij}(3\,\rm TeV),   
\nonumber
\\
&\tilde C_{2,dd}^{ij}(\Lambda_{\tt EW}) =  C_{2,dd}^{ji*}(\Lambda_{\tt EW}), \,
\tilde C_{3,dd}^{ij}(\Lambda_{\tt EW}) = C_{3,dd}^{ji*}(\Lambda_{\tt EW}),
\end{align}
where the numerical factors after the equality sign are the RG running effects from  
the NP scale $\Lambda_{\tt NP} =3\,\rm TeV$ to the EW scale $\Lambda_{\tt EW}=160\,\rm GeV$. 
By parameterizing $y_q^i \equiv |y_q^i| e^{i \alpha_i}$ and $y_d^i \equiv |y_d^i| e^{i \beta_i}$, we obtain from the LEFT master formula 
\begin{align}
\left| {2[M_{12}^{ij}]_{\tt NP}^{\tt LEFT} \over (\Delta M_{ij} )_{\tt exp} }\right| & =
\left| 
P_2^{ij} \left(C_{2,dd}^{ij} + \tilde  C_{2,dd}^{ij} \right)
+ P_3^{ij} \left(C_{3,dd}^{ij} + \tilde  C_{3,dd}^{ij} \right)
+ P_4^{ij}C_{4,dd}^{ij}
\right|(\Lambda_{\tt EW})
\nonumber
\\
& = { 10^{-6}\over {\rm TeV}^2 } 
\left| 
\left( 2.70 P_2^{ij}- 0.11 P_3^{ij}\right) 
\left(|y_q^i|^2 |y_d^j|^2 e^{ i \theta_{ij}} + ( |y_d^{i}|^2 |y_q^{j}|^2 e^{- i \theta_{ij}}\right)y_1^2 h_1(z)
\right.
\nonumber
\\
& \left. 
+ 3.74 P_4^{ij} |y_q^i| |y_q^{j}| |y_d^{i}|  |y_d^j| |y_1|^2 h_2(z)
\right|,
\end{align}    
where $\theta_{ij} \equiv \alpha_i + \alpha_j + \beta_i + \beta_j$.
Numerically, we fix $y_1=1$, $|y_q^s|=|y_d^s|=2$ as  in \cite{He:2024iju} and take real $y_{2}$ with $h_2 =h_1$, then 
\begin{subequations}
\begin{align}
\left| {2[M_{12}^{ds}]_{\tt NP}^{\tt LEFT} \over (\Delta M_{ds} )_{\tt exp} }\right| 
& =471
\left| 
|y_q^d|^2 e^{i \theta_{ds}}+ |y_d^{d}|^2 e^{-i \theta_{ds}} -4.49 |y_q^d| |y_d^{d}|
\right| |h_1|,
\\
\left| {2[M_{12}^{sb}]_{\tt NP}^{\tt LEFT} \over (\Delta M_{sb} )_{\tt exp} }\right| 
& = 0.24 
\left| 
|y_d^b|^2 e^{i \theta_{sb}} + |y_q^{b}|^2 e^{-i \theta_{sb}}
-3.20  |y_d^b| |y_q^{b}| 
\right| |h_1|, 
\\
\left| {2[M_{12}^{db}]_{\tt NP}^{\tt LEFT} \over (\Delta M_{db} )_{\tt exp} }\right| 
& = 1.36 \left| 
|y_q^d|^2 |y_d^b|^2 e^{i \theta_{db}}+ |y_d^{d}|^2 |y_q^{b}|^2 e^{-i \theta_{db}}
-3.50  |y_q^d||y_q^{b}| |y_d^{d}|  |y_d^b|
\right| |h_1|.
\end{align}
\end{subequations}
It can be seen that, for the $\calO(0.1)$ Yukawa couplings, the kaon mixing leads to
a stringent bound on $|h_1| \lesssim 10^{-3}$ to satisfy the requirement that the left-hand ratios be less than 0.1\,.

\begin{figure}[t]
\centering
\begin{tikzpicture}[mystyle,scale=1.4]
\begin{scope}[shift={(1,1)}]3
\draw[f,black] (-2,1) node[left]{$\texttt{q}$}  -- (-1,1);
\draw[f,magenta,ultra thick] (-1, 1) -- (0,1) node[midway,yshift=8pt]{$\texttt{Q}$};
\draw[f,black] (0, 1) -- (1,1) node[right]{$\texttt{q}$};
\draw[snar,black] (0, 1) -- (1,0.3) node[right]{$H$};
\draw[s,magenta,ultra thick] (-1,1) -- (-1,-1) node[midway,xshift = 8 pt]{$\Phi$};
\draw[f,black] (-1,-1) -- (-2,-1)node[left]{$\texttt{q}$};
\draw[f,magenta,ultra thick] (0,-1)--(-1, -1) node[midway,yshift=-8pt]{$\texttt{Q}$};
\draw[f,black] (1,-1) node[right]{$\texttt{q}$} -- (0, -1);
\draw[snar,black] (0, -1) -- (1,-0.3) node[right]{$H$};
\end{scope}
\end{tikzpicture}
\caption{Tree-level diagrams contributing to the dim-8 operators.}
\label{fig:dim8tree}
\end{figure}

The above example illustrates that the LO NP contribution to $\Delta F=2$ interactions arises first at dim-8 order through one-loop diagrams. 
For completeness, we briefly comment on possible NP scenarios with tree-level $\Delta F=2$ contributions at dim-8 order, 
while simultaneously suppressing or forbidding the tree-level dim-6 $\Delta F=2$ interactions.
We find that this can be easily realized in NP models that incorporate heavy vector-like  quarks and scalars,
similar to the aforementioned model, but without imposing an ad hoc discrete symmetry. Denoting a generic vector-like quark as $\texttt{Q}$ and a new scalar by $\Phi$, we anticipate the following Yukawa terms are allowed with an appropriate choice of $\rm SU(2)_L\otimes U(1)_Y$ representations for $\texttt{Q}$ and $\Phi$,
\begin{align}
{\cal L}_{\tt NP}\supset
\mathbb{Y}_1 \bar{\texttt{q}}_1 \texttt{Q} \Phi 
+ \mathbb{Y}_2 \bar{\texttt{q}}_2 \texttt{Q} H + \hc,
\end{align}
where $\texttt{q}_{1,2}$ represent the SM chiral quark fields $q,u,d$ and $\mathbb{Y}_{1,2}$ as new Yukawa couplings. By treating both $\texttt{Q}$ and $\Phi$ as heavy fields and integrating them out, we can generate the dim-8 four-quark-two-Higgs operators via the tree-level diagrams shown in \cref{fig:dim8tree}. 
For example, if we choose the SM representations, $\texttt{Q}\sim(\mathbf{3},\mathbf{1},-1/3)$ and 
$\Phi\sim(\mathbf{1},\mathbf{2},1/2)$, the corresponding new Yukawa terms are 
$\mathbb{Y}_1 \bar{q} \texttt{Q}_R \Phi 
+ \mathbb{Y}_2 \bar{q} \texttt{Q}_R H$. 
After integrating out $\texttt{Q}$ and $\Phi$, this model can yield operators $\calO_{q^4 H^2}^{(1-4)}$.

\section{Conclusions}
\label{sec:conclusions}

The $\Delta F=2$ neutral meson mixing processes are golden channels to probe NP parameter space due to their severe suppression in the standard model. In the LEFT framework there are eight operators describing each meson mixing. However, when going into the SMEFT framework, only four out of the eight operators can be generated at the dim-6 level by a tree-level matching, while the remaining four first appear at dimension 8 and were neglected in the previous analysis. Since many UV models generate these dim-8 instead of dim-6 operators at the leading order, it is important to consider their phenomenological consequences. In this work, we have studied the impact of these dim-8 operators on the meson mixing. We first collected the relevant dim-8 operators responsible for the neutral meson mixing and performed a tree-level matching with the eight LEFT operators at the electroweak scale. We then calculated the one-loop QCD renormalization group equations of these dim-8 operators. After that, we used the LEFT master formula to derive the bounds on the effective scale associated with those dim-8 operators. Surprisingly, we found the neutral meson mixing can probe the effective scale to tens of TeV, far beyond the sensitivity of other dim-8 operators to collider searches. Finally, we provided an example UV model to show the importance of these dim-8 operators for constraining NP scenarios.

\section*{Acknowledgements}
This work was supported by the Grants 
No.\,NSFC-12035008, 
No.\,NSFC-12247151, 
and No.\,NSFC-12305110.

\appendix
\section{The one-loop RGEs including gluon EoM contributions}
\label{app:RGE}

In this appendix we provide the full one-loop RGEs from the counterterms to the diagrams involving both internal and external gluons. For simplicity, we assume $N_c=3$ and $C_F=4/3$ in the anomalous dimension matrices directly. The operators $\calO_{q^2 \psi^2 H^2}^{(1, 2, 5,6)}$ with $\psi=u,d$ have no corrections to the gluon EoM diagrams due to their chiral and flavor structures. For $\calO_{q^2 \psi^2 H^2}^{(3,4)}$, the $\rm SU(2)_L$ singlet part $\bar \Psi \gamma_\mu T^A \Psi$ with $\Psi=q,u,d$ can contribute. For the Type-II operators, we have the RGEs 
\begin{subequations}
\begin{align}
\mu {d \over d\mu} C_{q^4 H^2}^{(2),xyzw} & = 
- {\alpha_s \over 2\pi }
\left[ C_{q^4 H^2}^{(2),xyzw}
-{3\over 2} C_{q^4 H^2}^{(2),xwzy}
-{3\over 2} C_{q^4 H^2}^{(2),zyxw}
- 6 C_{q^4 H^2}^{(5),xwzy}
\right.
\nonumber
\\
&\left. 
+ {1\over 18} \delta_{xy}\left( 2 C_{q^4 H^2}^{(2),zppw}+2 C_{q^4 H^2}^{(2),pwzp}
- 8 C_{q^4 H^2}^{(5),zppw}
+ C_{q^2 u^2 H^2}^{(4),zwpp}+ C_{q^2 d^2 H^2}^{(4),zwpp}
\right)
\right.
\nonumber
\\
&\left. 
- {1\over 12}\delta_{xw}\left( 
2 C_{q^4 H^2}^{(2),zppy} + 2 C_{q^4 H^2}^{(2),pyzp}
-8 C_{q^4 H^2}^{(5),zppy}
+ C_{q^2 u^2 H^2}^{(4),zypp}+ C_{q^2 d^2 H^2}^{(4),zypp}
\right)
\right.
\nonumber
\\
&\left. 
- {1\over 12}\delta_{yz}\left( 
2 C_{q^4 H^2}^{(2),xppw} + 2 C_{q^4 H^2}^{(2),pwxp}
-8 C_{q^4 H^2}^{(5),xppw}
+ C_{q^2 u^2 H^2}^{(4),xwpp}+ C_{q^2 d^2 H^2}^{(4),xwpp}
\right)
\right],
\\%
\mu {d \over d\mu} C_{q^4 H^2}^{(5),xyzw} & = - {\alpha_s \over 2\pi }
\left[
-{3\over 4} C_{q^4 H^2}^{(2),xwzy}
+{3\over 4} C_{q^4 H^2}^{(2),zyxw}
+ C_{q^4 H^2}^{(5),xyzw}
\right.
\nonumber
\\
&\left. 
- {1\over 24}\delta_{xw}\left( 
2 C_{q^4 H^2}^{(2),zppy} + 2 C_{q^4 H^2}^{(2),pyzp}
-8 C_{q^4 H^2}^{(5),zppy}
+ C_{q^2 u^2 H^2}^{(4),zypp}+ C_{q^2 d^2 H^2}^{(4),zypp}
\right)
\right.
\nonumber
\\
&\left. 
+ {1\over 24}\delta_{yz}\left( 
2 C_{q^4 H^2}^{(2),xppw} + 2 C_{q^4 H^2}^{(2),pwxp}
-8 C_{q^4 H^2}^{(5),xppw}
+ C_{q^2 u^2 H^2}^{(4),xwpp} + C_{q^2 d^2 H^2}^{(4),xwpp}
\right)
\right],
\\%
\mu {d \over d\mu} C_{q^2 \psi^2 H^2}^{(2),xyzw} & = - {\alpha_s \over 2\pi }
\left(
{4\over 3} C_{q^2 \psi^2 H^2}^{(4),xyzw}
\right),
\\
\mu {d \over d\mu} C_{q^2 \psi^2 H^2}^{(4),xyzw} & = - {\alpha_s \over 2\pi }
\left[
6 C_{q^2 \psi^2 H^2}^{(2),xyzw}
+7  C_{q^2 \psi^2 H^2}^{(4),xyzw}
\right.
\nonumber
\\
&\left.
- {1\over 3} \delta_{zw}\left( 
2 C_{q^4 H^2}^{(2),xppy} + 2 C_{q^4 H^2}^{(2),pyxp}
-8 C_{q^4 H^2}^{(5),xppy}
+ C_{q^2 u^2 H^2}^{(4),xypp}+ C_{q^2 d^2 H^2}^{(4),xypp}
\right)
\right],
\\%
\mu {d \over d\mu} C_{q^2 \psi^2 H^2}^{(5),xyzw} & = 
- {\alpha_s \over 2\pi }
\left( 8 C_{q^2 \psi^2 H^2}^{(5),xyzw}
- {32 \over 9} C_{q^2 \psi^2 H^2}^{(5),xwzy}
- {8 \over 9}C_{q^2 \psi^2 H^2}^{(6),xyzw}
- {56 \over 27} C_{q^2 \psi^2 H^2}^{(6),xyzw}
\right),
\\
\mu {d \over d\mu} C_{q^2 \psi^2 H^2}^{(6),xyzw} & =
- {\alpha_s \over 2\pi }
\left( -4  C_{q^2 \psi^2 H^2}^{(5),xyzw}
+ {8 \over 3} C_{q^2 \psi^2 H^2}^{(5),xwzy}
- {8 \over 3} C_{q^2 \psi^2 H^2}^{(6),xyzw}
- {22 \over 9} C_{q^2 \psi^2 H^2}^{(6),xyzw}
\right),
\end{align}
\end{subequations}
where the subscript $\psi=u,d$. The repeated superscript $p$ is implied to sum over the three generations of quarks, for instance, $C_{q^4 H^2}^{(2),zppw} =C_{q^4 H^2}^{(2),z11w} + C_{q^4 H^2}^{(2),z22w} +C_{q^4 H^2}^{(2),z33w}$, etc. 

For the Type-I operators, for the closure of RGEs, we have to include the other two operators 
\begin{subequations}
\begin{align}
\calO_{u^2 d^2 H^2}^{(1)} & = (\bar u \gamma_\mu u)(\bar d \gamma^\mu d)(H^\dagger H), 
\\
\calO_{u^2 d^2 H^2}^{(2)} & = (\bar u \gamma_\mu T^A u)(\bar d \gamma^\mu T^A d)(H^\dagger H), 
\end{align}
\end{subequations}
then the full RGEs are 
\begin{subequations}
\begin{align}
\mu {d \over d\mu} C_{q^4 H^2}^{(1),xyzw} & =
- {\alpha_s \over 2\pi} 
\left[ 
C_{q^4 H^2}^{(1),xyzw} 
- {3\over2} C_{q^4 H^2}^{(1),xwzy}
- {9\over2}  C_{q^4 H^2}^{(3),xwzy}
\right.
\nonumber
\\
&\left.
+ {1\over 36} \delta_{xy} \left( 4 C_{q^4 H^2}^{(1),zppw} 
+ 12 C_{q^4 H^2}^{(3),zppw}
+ C_{q^2 u^2 H^2}^{(3),zwpp}+ C_{q^2 d^2 H^2}^{(3),zwpp}
\right)
\right.
\nonumber
\\
&\left.
+ {1\over 36} \delta_{zw} \left( 4 C_{q^4 H^2}^{(1),xppy} 
+ 12 C_{q^4 H^2}^{(3),xppy}
+ C_{q^2 u^2 H^2}^{(3),xypp}+ C_{q^2 d^2 H^2}^{(3),xypp}
\right)
\right.
\nonumber
\\
&\left.
- {1\over 24} \delta_{xw} \left( 4 C_{q^4 H^2}^{(1),zppy} 
+ 12 C_{q^4 H^2}^{(3),zppy}
+ C_{q^2 u^2 H^2}^{(3),zypp}+ C_{q^2 d^2 H^2}^{(3),zypp}
\right)
\right.
\nonumber
\\
&\left.
-{1\over 24} \delta_{yz} \left( 4 C_{q^4 H^2}^{(1),xppw} 
+ 12 C_{q^4 H^2}^{(3),xppw}
+ C_{q^2 u^2 H^2}^{(3),xwpp}+ C_{q^2 d^2 H^2}^{(3),xwpp}
\right)
\right],
\\%
\mu {d \over d\mu} C_{q^4 H^2}^{(3),xyzw} & = 
- {\alpha_s \over 2\pi} 
\left[ 
- {3\over2}  C_{q^4 H^2}^{(1),xwzy}
+ C_{q^4 H^2}^{(3),xyzw}
+ {3\over2}  C_{q^4 H^2}^{(3),xyzw}
\right.
\nonumber
\\
&\left.
- {1\over 24} \delta_{xw} \left( 4 C_{q^4 H^2}^{(1),zppy} 
+ 12 C_{q^4 H^2}^{(3),zppy}
+ C_{q^2 u^2 H^2}^{(3),zypp}+ C_{q^2 d^2 H^2}^{(3),zypp}
\right)
\right.
\nonumber
\\
&\left.
-{1\over 24} \delta_{yz} \left( 4 C_{q^4 H^2}^{(1),xppw} 
+ 12 C_{q^4 H^2}^{(3),xppw}
+ C_{q^2 u^2 H^2}^{(3),xwpp}+ C_{q^2 d^2 H^2}^{(3),xwpp}
\right)
\right],
\\%
\mu {d \over d\mu} C_{q^2 \psi^2 H^2}^{(1),xyzw} & =
- {\alpha_s \over 2\pi} 
\left( {4\over3} C_{q^2 \psi^2 H^2}^{(1),xyzw} \right),
\\%
\mu {d \over d\mu} C_{q^2 \psi^2 H^2}^{(3),xyzw} & =
- {\alpha_s \over 2\pi} 
\left[ 
6 C_{q^2 \psi^2 H^2}^{(1),xyzw}
+ 7 C_{q^2 \psi^2 H^2}^{(3),xyzw}
\right.
\nonumber
\\
&\left. 
-{1\over3} \delta_{xy}
\left( 2 C_{q^2 \psi^2H^2}^{(3),ppzw} + 4 C_{\psi^4 H^2}^{zppw} 
+ C_{u^2 d^2 H^2}^{(2), zwpp[ppzw]}\right)
\right.
\nonumber
\\
&\left. 
-{1\over3} \delta_{zw}
\left( 4 C_{q^4 H^2}^{(1),xppy} 
+ 12 C_{q^4 H^2}^{(3),xppy}
+ C_{q^2 u^2 H^2}^{(3),xypp}+ C_{q^2 d^2 H^2}^{(3),xypp} \right)
\right],
\\%
\mu {d \over d\mu} C_{\psi^4 H^2}^{xyzw} & =
- {\alpha_s \over 2\pi} 
\left[ 
C_{\psi^4 H^2}^{xyzw} 
-3 C_{\psi^4 H^2}^{xwzy}
\right.
\nonumber
\\
&\left.
+ {1\over 36} \delta_{xy} \left( 2 C_{q^2 \psi^2 H^2}^{(3),ppzw} 
+ 4 C_{\psi^4 H^2}^{zppw} 
+ C_{u^2 d^2 H^2}^{(2), zwpp[ppzw]}
\right)
\right.
\nonumber
\\
&\left.
+ {1\over 36} \delta_{zw} \left(2 C_{q^2 \psi^2 H^2}^{(3),ppxy} 
+ 4 C_{\psi^4 H^2}^{xppy} 
+ C_{u^2 d^2 H^2}^{(2), xypp[ppxy]}
\right)
\right.
\nonumber
\\
&\left.
- {1\over 12} \delta_{xw} \left( 2 C_{q^2 \psi^2 H^2}^{(3),ppzy} 
+ 4 C_{\psi^4 H^2}^{zppy} 
+ C_{u^2 d^2 H^2}^{(2), zypp [ppzy]}
\right)
\right.
\nonumber
\\
&\left.
-{1\over 12} \delta_{yz} \left( 2 C_{q^2 \psi^2 H^2}^{(3),ppxw} 
+ 4 C_{\psi^4 H^2}^{xppw} 
+ C_{u^2 d^2 H^2}^{(2), xwpp[ppxw]}
\right)
\right],
\\%
\mu {d \over d\mu} C_{u^2 d^2 H^2}^{(1),xyzw} & =
- {\alpha_s \over 2\pi} 
\left( - {4\over3}  C_{u^2 d^2 H^2}^{(1),xyzw} \right),
\\
\mu {d \over d\mu} C_{u^2 d^2 H^2}^{(2),xyzw} & =
- {\alpha_s \over 2\pi} 
\left[ 
- 6 C_{u^2 d^2 H^2}^{(1),xyzw}
+ 2 C_{u^2 d^2 H^2}^{(2),xyzw}
\right.
\nonumber
\\
&\left.
- {1\over 3} \delta_{xy} \left( 2 C_{q^2 d^2 H^2}^{(3),ppzw} 
+ 4 C_{d^4 H^2}^{zppw} 
+ C_{u^2 d^2 H^2}^{(2),ppzw}
\right)
\right.
\nonumber
\\
&\left.
- {1\over 3} \delta_{zw} \left(2 C_{q^2 u^2 H^2}^{(3),ppxy} 
+ 4 C_{u^4 H^2}^{xppy} 
+ C_{u^2 d^2 H^2}^{(2),xypp}
\right)
\right],
\end{align}
\end{subequations}
where again the subscript $\psi=u,d$. For the WC $C_{u^2 d^2 H^2}^{(2)}$ appearing in the RGEs of $C_{q^2 \psi^2 H^2}^{(3),xyzw}$ and $C_{\psi^4 H^2}^{xyzw}$, 
the indices outside (inside) the square brackets are for $\psi=u$ ($d$), respectively.  
From the above results, one finds the additional contributions due to the gluon EoM operators always appear with a Kronecker delta. 
The RGEs for the Type-I operators are consistent with those of the corresponding dim-6 operators without the $H^\dagger H$ factor given in \cite{Alonso:2013hga}.

\section{The results in down-quark flavor basis}
\label{app:downbasis}

\begin{table}[h]
\center
\resizebox{\linewidth}{!}{
\renewcommand\arraystretch{1.7}
\begin{tabular}{| c | l |}
\hline
\cellcolor{magenta!25}LEFT operators & \multicolumn{1}{|c|}{\cellcolor{magenta!25}Matching results at electroweak scale $\Lambda_{\tt EW}$ in the down-flavor basis}
\\
\hline
\multirow{2}*{$\calO_{1}^{ij}= (\bar{q}_{i}^\alpha \gamma_\mu P_L q_{j}^\alpha) (\bar{q}_{i}^\beta \gamma^\mu P_L q_{j}^\beta)$}
 & $C_{1,dd}^{ij}=
 { v^2\over 2}  
 \left( C_{q^4 H^2}^{(2),ijij} 
 {\color{cyan} + C_{q^4 H^2}^{(1),ijij} + C_{q^4 H^2}^{(3),ijij} }
 \right) 
{\color{blue} + C_{qq}^{(1),ijij} + C_{qq}^{(3),ijij} }$
\\
\cline{2-2}
& $C_{1,uu}^{ij}= 
\left[  -{ v^2\over 2} \left(  C_{q^4 H^2}^{(2),xyzw} {\color{cyan} - C_{q^4 H^2}^{(1),xyzw} - C_{q^4 H^2}^{(3),xyzw} } \right)
{\color{blue} + C_{qq}^{(1),xyzw} + C_{qq}^{(3),xyzw}}  \right] V_{ix}V_{jy}^* V_{iz} V_{jw}^*$
\\
\hline%
\multirow{2}*{$\calO_{2}^{ij}= (\bar{q}_{i}^\alpha P_L q_{j}^\alpha) (\bar{q}_{i}^\beta P_L q_{j}^\beta)$} 
& $C^{ij}_{2,dd}= {v^2 \over 2}\left(C^{(5),jiji*}_{q^2 d^2 H^2}-{1\over 6}C^{(6),jiji*}_{q^2 d^2 H^2}  \right) $
\\
\cline{2-2}
 & $C^{ij}_{2,uu}= {v^2 \over 2} 
 \left( C^{(5),xizi*}_{q^2 u^2 H^2} -{1 \over 6} C^{(6),xizi*}_{q^2 u^2 H^2} \right) V_{jx}^* V_{jz}^* $
\\
\hline%
\multirow{2}*{$\calO_{3}^{ij} = (\bar{q}_{i}^\alpha P_L q_{j}^\beta) (\bar{q}_{i}^\beta P_L q_{j}^\alpha)$} 
& $C_{3,dd}^{ij} = {v^2 \over 4} C^{(6),jiji *}_{q^2 d^2 H^2} $
\\
\cline{2-2}
 & $C_{3,uu}^{ij} = {v^2 \over 4} C^{(6),xizi *}_{q^2 u^2 H^2} V_{jx}^* V_{jz}^* $
 \\
\hline%
\multirow{2}*{$\calO_{4}^{ij} = (\bar{q}_{i}^\alpha P_L q_{j}^\alpha) (\bar{q}_{i}^\beta P_R q_{j}^\beta)$} 
& $C_{4,dd}^{ij} = 
- {v^2 \over 2} \left( C^{(4),ijij}_{q^2 d^2 H^2} 
{\color{cyan} + C_{q^2 d^2 H^2}^{(3),ijij} }  \right) 
{\color{blue}-C_{qd}^{(8),ijij}}$
\\
\cline{2-2}
 & $C_{4,uu}^{ij}=\left[  {v^2 \over 2} \left( 
 C^{(4),xyij}_{q^2 u^2 H^2} {\color{cyan} - C^{(3),xyij}_{q^2 u^2 H^2} }
 \right)
 {\color{blue}-C_{qu}^{(8),xyij}}  \right]
V_{ix} V_{jy}^*  $
\\
\hline
\multirow{2}*{$\calO_{5}^{ij} = (\bar{q}_{i}^\alpha P_L q_{j}^\beta) (\bar{q}_{i}^\beta P_R q_{j}^\alpha)$} & 
$C_{5,dd}^{ij} = 
 - v^2 \left( 
C^{(2),ijij}_{q^2 d^2 H^2} 
- {1\over 6} C^{(4),ijij}_{q^2 d^2 H^2}  
{\color{cyan}+C_{q^2 d^2 H^2}^{(1),ijij} 
- {1\over 6 } C_{q^2 d^2 H^2}^{(3),ijij} } \right) 
{\color{blue}- 2 C_{qd}^{(1),ijij} + {1\over 3 }C_{qd}^{(8),ijij}}
$
\\
\cline{2-2}
 & $C_{5,uu}^{ij}=\left[ v^2 \left( C^{(2),xyij}_{q^2 u^2 H^2}
 -{1\over 6} C^{(4),xyij}_{q^2 u^2 H^2} {\color{cyan}- C_{q^2 u^2 H^2}^{(1),xyij}+{1\over 6} C^{(3),xyij}_{q^2 u^2 H^2} } \right) 
 {\color{blue} -2 C_{qu}^{(1),xyij} + {1\over3}C_{qu}^{(8),xyij}} \right] V_{ix} V_{jy}^*$
\\
\hline
\end{tabular} 
}
\caption{The $\Delta F=2$ tree-level matching results in the down-flavor basis. The results for $\tilde\calO_1^{ij}$ are identical to those provided in \cref{tab:matching} and are not repeated here. }
\label{tab:matchingdownbasis}
\end{table}

\begin{figure}
\centering
\includegraphics[width=0.49\linewidth]{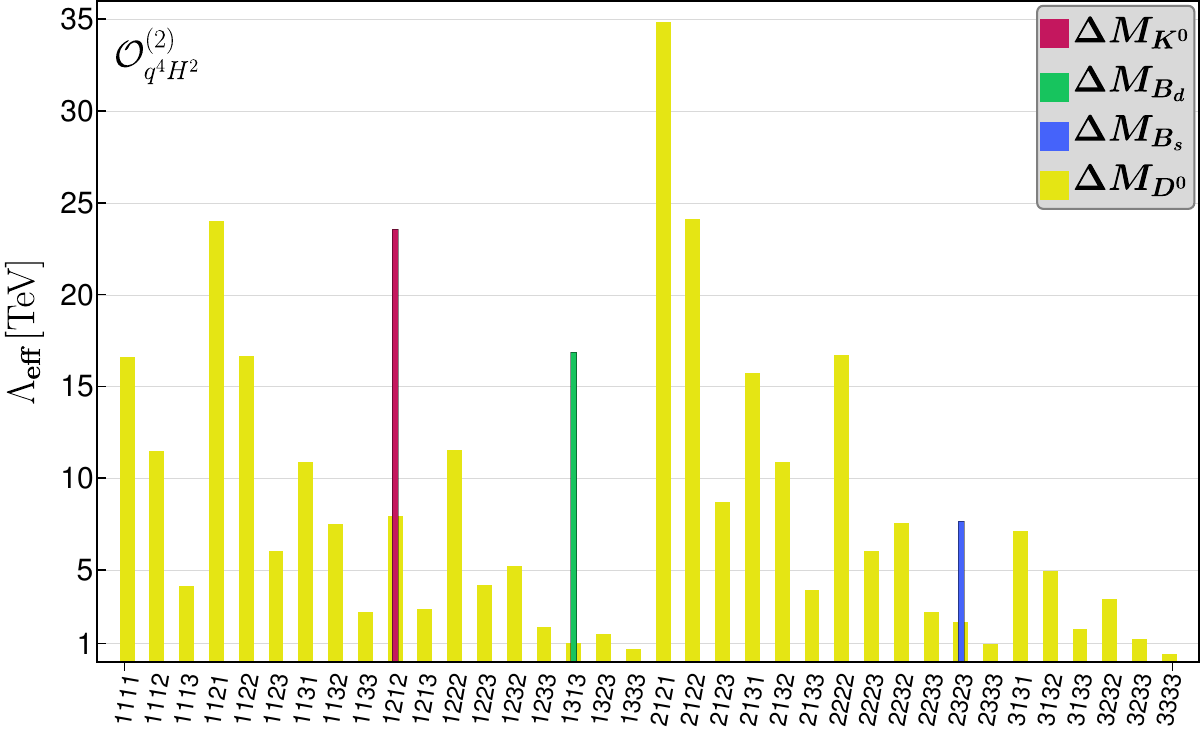}
\includegraphics[width=0.495\linewidth]{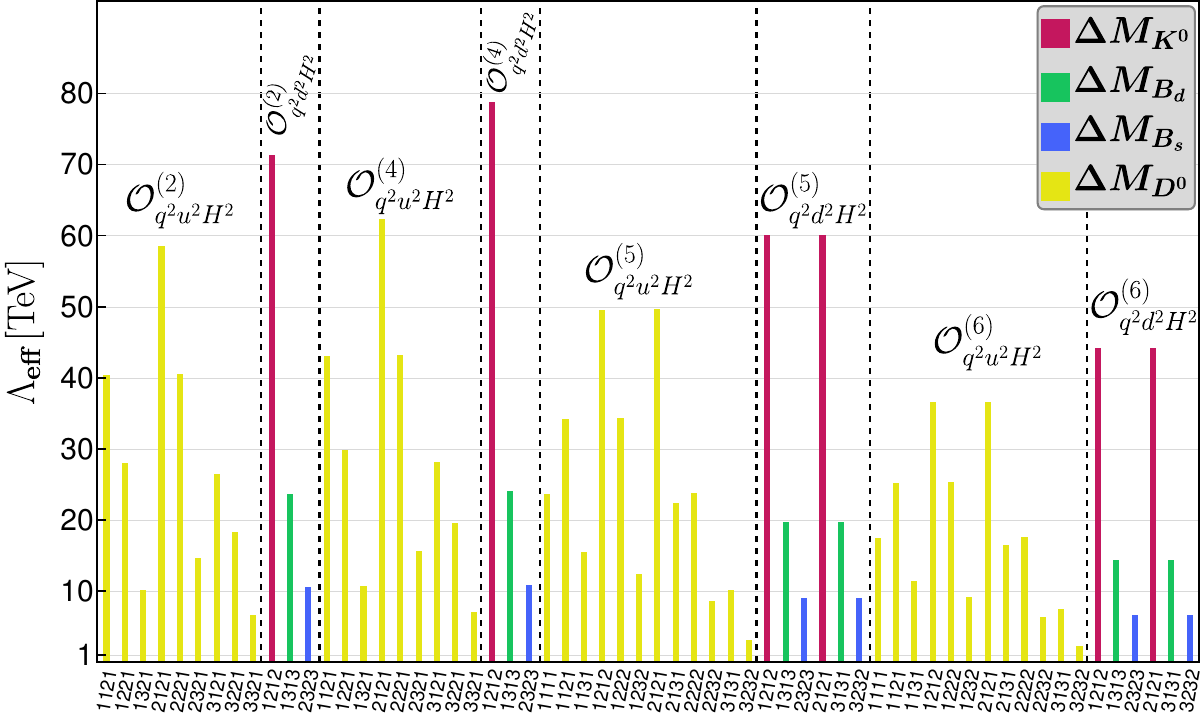}
\vspace{0.1cm}
\\
\includegraphics[width=0.49\linewidth]{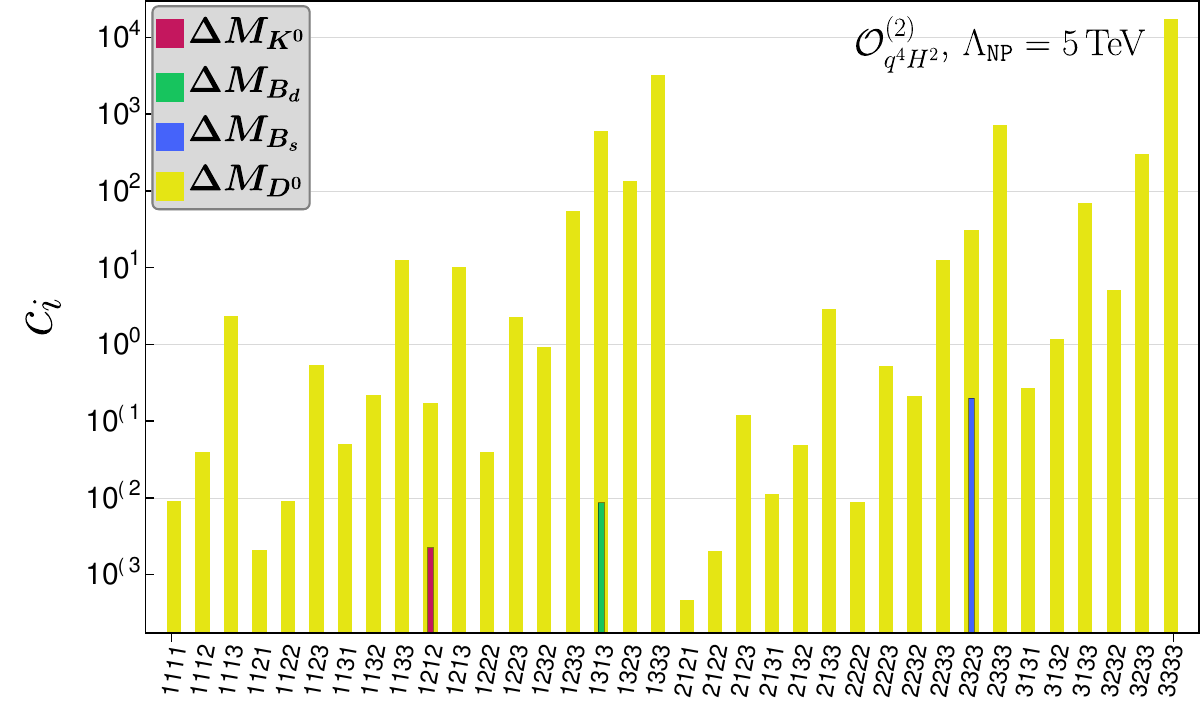}
\includegraphics[width=0.495\linewidth]{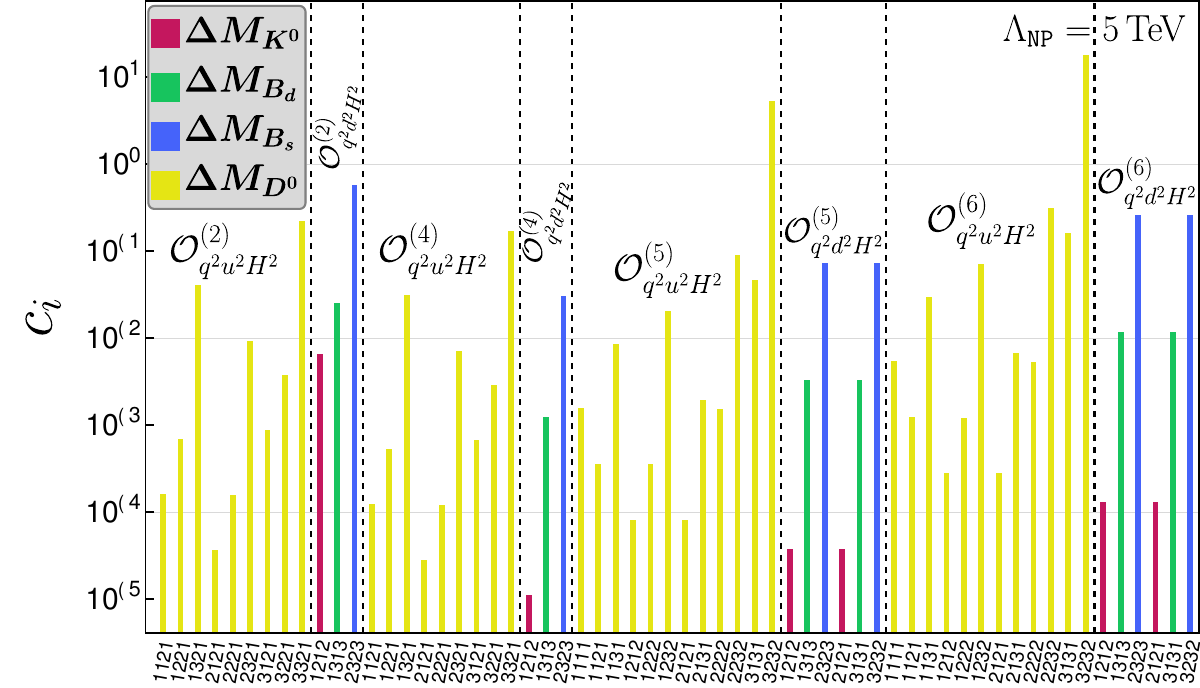}
\caption{Constraints in the down-quark flavor basis. 
The upper panels show the lower bound on the effective scale $\Lambda_{\tt eff}$, while the lower panels provide the upper bounds on the dimensionless WCs $c_i$ at $\mu=\Lambda_{\tt NP}=5~\textrm{TeV}$ after taking into account the RGE effects.}
\label{fig:downbasis_results}
\end{figure}

While phenomenological analyses of UV models may be performed in either the up-quark or down-quark flavor basis, presenting results in both bases will facilitate comparing 
constraints on the parameters of UV models. 
We discuss our results now in the down-quark flavor basis where the CKM matrix elements are associated with the left-handed up-type quarks. Our matching results are given in \cref{tab:matchingdownbasis}, which should be compared with those in the up-quark flavor basis provided in \cref{tab:matching}. Based on the matching results, we present the numerical results in \cref{fig:downbasis_results}, including constraints on the effective scale $\Lambda_{\tt eff}$ associated with each dim-8 operator, and the dimensionless WCs $c_i$ at $\mu=\Lambda_{\tt NP}=5$ TeV with RGE effects.

As can be seen from comparison between \cref{fig:downbasis_results} and \cref{fig:cons-noRGE,fig:cc-RGE}, the results differ from one another due to different placements of the CKM matrix in the two bases, so that even the explored flavor structure differs for each given operator. In the down-quark basis, more flavor components can be constrained through the $D^0$ meson mixing, while the constraints from $\Delta K^0$ and $\Delta B_{d,s}$ are more specific. The situation is reversed in the up-quark basis, with $\Delta K^0$ and $\Delta B_{d,s}$ providing broader constraints. 
On the other hand, the equal-constraint relationship between Type-I and Type-II operators in \cref{sec:numerical} remains unchanged in the treatment of one operator a time, even after incorporating the QCD RGE effects. 
Moreover, the results for the pure right-handed operators $\calO_{u^4 H^2}$ and $\calO_{d^4 H^2}$ are identical in the two bases. 
As a result, the constraints for Type-I operators are not separately shown here.

\bibliography{references.bib}{}

\providecommand{\href}[2]{#2}\begingroup\raggedright\begin{thebibliography}{10}

\bibitem{KOTO:2018dsc}
{\scshape KOTO} collaboration, J.~K. Ahn et~al., \emph{{Search for the $K_L
  \!\to\! \pi^0 \nu \overline{\nu}$ and $K_L \!\to\! \pi^0 X^0$ decays at the
  J-PARC KOTO experiment}},
  \href{http://dx.doi.org/10.1103/PhysRevLett.122.021802}{\emph{Phys. Rev.
  Lett.} {\bf 122} (2019) 021802}, [\href{http://arxiv.org/abs/1810.09655}{{\tt
  1810.09655}}].

\bibitem{Belle-II:2023esi}
{\scshape Belle-II} collaboration, I.~Adachi et~al., \emph{{Evidence for
  B+\textrightarrow{}K+\ensuremath{\nu}\ensuremath{\nu}\textasciimacron{}
  decays}}, \href{http://dx.doi.org/10.1103/PhysRevD.109.112006}{\emph{Phys.
  Rev. D} {\bf 109} (2024) 112006},
  [\href{http://arxiv.org/abs/2311.14647}{{\tt 2311.14647}}].

\bibitem{Buchalla:1995vs}
G.~Buchalla, A.~J. Buras and M.~E. Lautenbacher, \emph{{Weak decays beyond
  leading logarithms}},
  \href{http://dx.doi.org/10.1103/RevModPhys.68.1125}{\emph{Rev. Mod. Phys.}
  {\bf 68} (1996) 1125--1144}, [\href{http://arxiv.org/abs/hep-ph/9512380}{{\tt
  hep-ph/9512380}}].

\bibitem{Jenkins:2017jig}
E.~E. Jenkins, A.~V. Manohar and P.~Stoffer, \emph{{Low-Energy Effective Field
  Theory below the Electroweak Scale: Operators and Matching}},
  \href{http://dx.doi.org/10.1007/JHEP03(2018)016}{\emph{JHEP} {\bf 03} (2018)
  016}, [\href{http://arxiv.org/abs/1709.04486}{{\tt 1709.04486}}].

\bibitem{Liao:2020zyx}
Y.~Liao, X.-D. Ma and Q.-Y. Wang, \emph{{Extending low energy effective field
  theory with a complete set of dimension-7 operators}},
  \href{http://dx.doi.org/10.1007/JHEP08(2020)162}{\emph{JHEP} {\bf 08} (2020)
  162}, [\href{http://arxiv.org/abs/2005.08013}{{\tt 2005.08013}}].

\bibitem{Grzadkowski:2010es}
B.~Grzadkowski, M.~Iskrzynski, M.~Misiak and J.~Rosiek, \emph{{Dimension-Six
  Terms in the Standard Model Lagrangian}},
  \href{http://dx.doi.org/10.1007/JHEP10(2010)085}{\emph{JHEP} {\bf 10} (2010)
  085}, [\href{http://arxiv.org/abs/1008.4884}{{\tt 1008.4884}}].

\bibitem{Lehman:2014jma}
L.~Lehman, \emph{{Extending the Standard Model Effective Field Theory with the
  Complete Set of Dimension-7 Operators}},
  \href{http://dx.doi.org/10.1103/PhysRevD.90.125023}{\emph{Phys. Rev. D} {\bf
  90} (2014) 125023}, [\href{http://arxiv.org/abs/1410.4193}{{\tt 1410.4193}}].

\bibitem{Murphy:2020rsh}
C.~W. Murphy, \emph{{Dimension-8 operators in the Standard Model Eective Field
  Theory}}, \href{http://dx.doi.org/10.1007/JHEP10(2020)174}{\emph{JHEP} {\bf
  10} (2020) 174}, [\href{http://arxiv.org/abs/2005.00059}{{\tt 2005.00059}}].

\bibitem{Liao:2016hru}
Y.~Liao and X.-D. Ma, \emph{{Renormalization Group Evolution of Dimension-seven
  Baryon- and Lepton-number-violating Operators}},
  \href{http://dx.doi.org/10.1007/JHEP11(2016)043}{\emph{JHEP} {\bf 11} (2016)
  043}, [\href{http://arxiv.org/abs/1607.07309}{{\tt 1607.07309}}].

\bibitem{Li:2020gnx}
H.-L. Li, Z.~Ren, J.~Shu, M.-L. Xiao, J.-H. Yu and Y.-H. Zheng, \emph{{Complete
  set of dimension-eight operators in the standard model effective field
  theory}}, \href{http://dx.doi.org/10.1103/PhysRevD.104.015026}{\emph{Phys.
  Rev. D} {\bf 104} (2021) 015026},
  [\href{http://arxiv.org/abs/2005.00008}{{\tt 2005.00008}}].

\bibitem{Liao:2020jmn}
Y.~Liao and X.-D. Ma, \emph{{An explicit construction of the dimension-9
  operator basis in the standard model effective field theory}},
  \href{http://dx.doi.org/10.1007/JHEP11(2020)152}{\emph{JHEP} {\bf 11} (2020)
  152}, [\href{http://arxiv.org/abs/2007.08125}{{\tt 2007.08125}}].

\bibitem{Li:2020xlh}
H.-L. Li, Z.~Ren, M.-L. Xiao, J.-H. Yu and Y.-H. Zheng, \emph{{Complete set of
  dimension-nine operators in the standard model effective field theory}},
  \href{http://dx.doi.org/10.1103/PhysRevD.104.015025}{\emph{Phys. Rev. D} {\bf
  104} (2021) 015025}, [\href{http://arxiv.org/abs/2007.07899}{{\tt
  2007.07899}}].

\bibitem{Harlander:2023psl}
R.~V. Harlander, T.~Kempkens and M.~C. Schaaf, \emph{{Standard model effective
  field theory up to mass dimension 12}},
  \href{http://dx.doi.org/10.1103/PhysRevD.108.055020}{\emph{Phys. Rev. D} {\bf
  108} (2023) 055020}, [\href{http://arxiv.org/abs/2305.06832}{{\tt
  2305.06832}}].

\bibitem{Aebischer:2020dsw}
J.~Aebischer, C.~Bobeth, A.~J. Buras and J.~Kumar, \emph{{SMEFT ATLAS of
  $\Delta$F = 2 transitions}},
  \href{http://dx.doi.org/10.1007/JHEP12(2020)187}{\emph{JHEP} {\bf 12} (2020)
  187}, [\href{http://arxiv.org/abs/2009.07276}{{\tt 2009.07276}}].

\bibitem{Aebischer:2022anv}
J.~Aebischer, A.~J. Buras and J.~Kumar, \emph{{NLO QCD renormalization group
  evolution for nonleptonic \ensuremath{\Delta}F=2 transitions in the SMEFT}},
  \href{http://dx.doi.org/10.1103/PhysRevD.106.035003}{\emph{Phys. Rev. D} {\bf
  106} (2022) 035003}, [\href{http://arxiv.org/abs/2203.11224}{{\tt
  2203.11224}}].

\bibitem{ValeSilva:2022tph}
L.~Vale~Silva, \emph{{Effects of squared four-fermion operators of the standard
  model effective field theory on meson mixing}},
  \href{http://dx.doi.org/10.1103/PhysRevD.110.016006}{\emph{Phys. Rev. D} {\bf
  110} (2024) 016006}, [\href{http://arxiv.org/abs/2201.03038}{{\tt
  2201.03038}}].

\bibitem{Bhattacharya:2023beo}
S.~Bhattacharya, S.~Jahedi, S.~Nandi and A.~Sarkar, \emph{{Probing flavor
  constrained SMEFT operators through tc production at the muon collider}},
  \href{http://dx.doi.org/10.1007/JHEP07(2024)061}{\emph{JHEP} {\bf 07} (2024)
  061}, [\href{http://arxiv.org/abs/2312.14872}{{\tt 2312.14872}}].

\bibitem{He:2024iju}
X.-G. He, X.-D. Ma, M.~A. Schmidt, G.~Valencia and R.~R. Volkas, \emph{{Scalar
  dark matter explanation of the excess in the Belle II
  B$^{+}$\textrightarrow{} K$^{+}$+ invisible measurement}},
  \href{http://dx.doi.org/10.1007/JHEP07(2024)168}{\emph{JHEP} {\bf 07} (2024)
  168}, [\href{http://arxiv.org/abs/2403.12485}{{\tt 2403.12485}}].

\bibitem{Liu:2016idz}
D.~Liu, A.~Pomarol, R.~Rattazzi and F.~Riva, \emph{{Patterns of Strong Coupling
  for LHC Searches}},
  \href{http://dx.doi.org/10.1007/JHEP11(2016)141}{\emph{JHEP} {\bf 11} (2016)
  141}, [\href{http://arxiv.org/abs/1603.03064}{{\tt 1603.03064}}].

\bibitem{Contino:2016jqw}
R.~Contino, A.~Falkowski, F.~Goertz, C.~Grojean and F.~Riva, \emph{{On the
  Validity of the Effective Field Theory Approach to SM Precision Tests}},
  \href{http://dx.doi.org/10.1007/JHEP07(2016)144}{\emph{JHEP} {\bf 07} (2016)
  144}, [\href{http://arxiv.org/abs/1604.06444}{{\tt 1604.06444}}].

\bibitem{Dawson:2022cmu}
S.~Dawson, D.~Fontes, S.~Homiller and M.~Sullivan, \emph{{Role of
  dimension-eight operators in an EFT for the 2HDM}},
  \href{http://dx.doi.org/10.1103/PhysRevD.106.055012}{\emph{Phys. Rev. D} {\bf
  106} (2022) 055012}, [\href{http://arxiv.org/abs/2205.01561}{{\tt
  2205.01561}}].

\bibitem{Gabbiani:1996hi}
F.~Gabbiani, E.~Gabrielli, A.~Masiero and L.~Silvestrini, \emph{{A Complete
  analysis of FCNC and CP constraints in general SUSY extensions of the
  standard model}},
  \href{http://dx.doi.org/10.1016/0550-3213(96)00390-2}{\emph{Nucl. Phys. B}
  {\bf 477} (1996) 321--352}, [\href{http://arxiv.org/abs/hep-ph/9604387}{{\tt
  hep-ph/9604387}}].

\bibitem{FLAG:2019iem}
{\scshape Flavour Lattice Averaging Group} collaboration, S.~Aoki et~al.,
  \emph{{FLAG Review 2019: Flavour Lattice Averaging Group (FLAG)}},
  \href{http://dx.doi.org/10.1140/epjc/s10052-019-7354-7}{\emph{Eur. Phys. J.
  C} {\bf 80} (2020) 113}, [\href{http://arxiv.org/abs/1902.08191}{{\tt
  1902.08191}}].

\bibitem{Buras:2000if}
A.~J. Buras, M.~Misiak and J.~Urban, \emph{{Two loop QCD anomalous dimensions
  of flavor changing four quark operators within and beyond the standard
  model}}, \href{http://dx.doi.org/10.1016/S0550-3213(00)00437-5}{\emph{Nucl.
  Phys. B} {\bf 586} (2000) 397--426},
  [\href{http://arxiv.org/abs/hep-ph/0005183}{{\tt hep-ph/0005183}}].

\bibitem{Buras:2022cyc}
A.~J. Buras, \emph{{$\varepsilon'/\varepsilon$ in the Standard Model and
  Beyond: 2021}},  in \emph{{11th International Workshop on the CKM Unitarity
  Triangle}}, 3, 2022.
\newblock \href{http://arxiv.org/abs/2203.12632}{{\tt 2203.12632}}.

\bibitem{Hamoudou:2022tdn}
S.~Hamoudou, J.~Kumar and D.~London, \emph{{Dimension-8 SMEFT matching
  conditions for the low-energy effective field theory}},
  \href{http://dx.doi.org/10.1007/JHEP03(2023)157}{\emph{JHEP} {\bf 03} (2023)
  157}, [\href{http://arxiv.org/abs/2207.08856}{{\tt 2207.08856}}].

\bibitem{Harnik:2012pb}
R.~Harnik, J.~Kopp and J.~Zupan, \emph{{Flavor Violating Higgs Decays}},
  \href{http://dx.doi.org/10.1007/JHEP03(2013)026}{\emph{JHEP} {\bf 03} (2013)
  026}, [\href{http://arxiv.org/abs/1209.1397}{{\tt 1209.1397}}].

\bibitem{Liao:2019tep}
Y.~Liao and X.-D. Ma, \emph{{Renormalization Group Evolution of Dimension-seven
  Operators in Standard Model Effective Field Theory and Relevant
  Phenomenology}}, \href{http://dx.doi.org/10.1007/JHEP03(2019)179}{\emph{JHEP}
  {\bf 03} (2019) 179}, [\href{http://arxiv.org/abs/1901.10302}{{\tt
  1901.10302}}].

\bibitem{Liao:2019gex}
Y.~Liao, X.-D. Ma and H.-L. Wang, \emph{{Effective field theory approach to
  lepton number violating decays $K^\pm\rightarrow \pi^\mp l^{\pm}l^{\pm}$:
  short-distance contribution}},
  \href{http://dx.doi.org/10.1007/JHEP01(2020)127}{\emph{JHEP} {\bf 01} (2020)
  127}, [\href{http://arxiv.org/abs/1909.06272}{{\tt 1909.06272}}].

\bibitem{Wang:2022lfq}
B.~Wang, \emph{{Calculating $\Delta m_K$ with lattice QCD}},
  \href{http://dx.doi.org/10.22323/1.396.0141}{\emph{PoS} {\bf LATTICE2021}
  (2022) 141}, [\href{http://arxiv.org/abs/2301.01387}{{\tt 2301.01387}}].

\bibitem{ParticleDataGroup:2024cfk}
{\scshape Particle Data Group} collaboration, S.~Navas et~al., \emph{{Review of
  particle physics}},
  \href{http://dx.doi.org/10.1103/PhysRevD.110.030001}{\emph{Phys. Rev. D} {\bf
  110} (2024) 030001}.

\bibitem{HFLAV:2022esi}
{\scshape HFLAV} collaboration, Y.~S. Amhis et~al., \emph{{Averages of
  b-hadron, c-hadron, and \ensuremath{\tau}-lepton properties as of 2021}},
  \href{http://dx.doi.org/10.1103/PhysRevD.107.052008}{\emph{Phys. Rev. D} {\bf
  107} (2023) 052008}, [\href{http://arxiv.org/abs/2206.07501}{{\tt
  2206.07501}}].

\bibitem{Albrecht:2024oyn}
J.~Albrecht, F.~Bernlochner, A.~Lenz and A.~Rusov, \emph{{Lifetimes of
  b-hadrons and mixing of neutral B-mesons: theoretical and experimental
  status}}, \href{http://dx.doi.org/10.1140/epjs/s11734-024-01124-3}{\emph{Eur.
  Phys. J. ST} {\bf 233} (2024) 359--390},
  [\href{http://arxiv.org/abs/2402.04224}{{\tt 2402.04224}}].

\bibitem{Lenz:2019lvd}
A.~Lenz and G.~Tetlalmatzi-Xolocotzi, \emph{{Model-independent bounds on new
  physics effects in non-leptonic tree-level decays of B-mesons}},
  \href{http://dx.doi.org/10.1007/JHEP07(2020)177}{\emph{JHEP} {\bf 07} (2020)
  177}, [\href{http://arxiv.org/abs/1912.07621}{{\tt 1912.07621}}].

\bibitem{DiLuzio:2019jyq}
L.~Di~Luzio, M.~Kirk, A.~Lenz and T.~Rauh, \emph{{$\Delta M_s$ theory precision
  confronts flavour anomalies}},
  \href{http://dx.doi.org/10.1007/JHEP12(2019)009}{\emph{JHEP} {\bf 12} (2019)
  009}, [\href{http://arxiv.org/abs/1909.11087}{{\tt 1909.11087}}].

\bibitem{Nierste:2009wg}
U.~Nierste, \emph{{Three Lectures on Meson Mixing and CKM phenomenology}},  in
  \emph{{Helmholz International Summer School on Heavy Quark Physics}},
  pp.~1--38, 3, 2009.
\newblock \href{http://arxiv.org/abs/0904.1869}{{\tt 0904.1869}}.

\bibitem{Nelson:1999fg}
H.~N. Nelson, \emph{{Compilation of D0 ---\ensuremath{>} anti-D0 mixing
  predictions}},  in \emph{{19th International Symposium on Lepton and Photon
  Interactions at High-Energies (LP 99)}}, 8, 1999.
\newblock \href{http://arxiv.org/abs/hep-ex/9908021}{{\tt hep-ex/9908021}}.

\bibitem{Petrov:2003un}
A.~A. Petrov, \emph{{Charm physics: Theoretical review}}, {\emph{eConf} {\bf
  C030603} (2003) MEC05}, [\href{http://arxiv.org/abs/hep-ph/0311371}{{\tt
  hep-ph/0311371}}].

\bibitem{Chavez:2012xt}
C.~A. Chavez, R.~F. Cowan and W.~S. Lockman, \emph{{Charm meson mixing: An
  experimental review}},
  \href{http://dx.doi.org/10.1142/S0217751X12300190}{\emph{Int. J. Mod. Phys.
  A} {\bf 27} (2012) 1230019}, [\href{http://arxiv.org/abs/1209.5806}{{\tt
  1209.5806}}].

\bibitem{Lenz:2020awd}
A.~Lenz and G.~Wilkinson, \emph{{Mixing and CP Violation in the Charm System}},
  \href{http://dx.doi.org/10.1146/annurev-nucl-102419-124613}{\emph{Ann. Rev.
  Nucl. Part. Sci.} {\bf 71} (2021) 59--85},
  [\href{http://arxiv.org/abs/2011.04443}{{\tt 2011.04443}}].

\bibitem{Golowich:2007ka}
E.~Golowich, J.~Hewett, S.~Pakvasa and A.~A. Petrov, \emph{{Implications of
  $D^0$ - $\bar{D}^0$ Mixing for New Physics}},
  \href{http://dx.doi.org/10.1103/PhysRevD.76.095009}{\emph{Phys. Rev. D} {\bf
  76} (2007) 095009}, [\href{http://arxiv.org/abs/0705.3650}{{\tt 0705.3650}}].

\bibitem{Bazavov:2017weg}
A.~Bazavov et~al., \emph{{Short-distance matrix elements for $D^0$-meson mixing
  for $N_f=2+1$ lattice QCD}},
  \href{http://dx.doi.org/10.1103/PhysRevD.97.034513}{\emph{Phys. Rev. D} {\bf
  97} (2018) 034513}, [\href{http://arxiv.org/abs/1706.04622}{{\tt
  1706.04622}}].

\bibitem{Boyle:2017ssm}
P.~Boyle, N.~Garron, J.~Kettle, A.~Khamseh and J.~T. Tsang, \emph{{BSM Kaon
  Mixing at the Physical Point}},
  \href{http://dx.doi.org/10.1051/epjconf/201817513010}{\emph{EPJ Web Conf.}
  {\bf 175} (2018) 13010}, [\href{http://arxiv.org/abs/1710.09176}{{\tt
  1710.09176}}].

\bibitem{FermilabLattice:2016ipl}
{\scshape Fermilab Lattice, MILC} collaboration, A.~Bazavov et~al.,
  \emph{{$B^0_{(s)}$-mixing matrix elements from lattice QCD for the Standard
  Model and beyond}},
  \href{http://dx.doi.org/10.1103/PhysRevD.93.113016}{\emph{Phys. Rev. D} {\bf
  93} (2016) 113016}, [\href{http://arxiv.org/abs/1602.03560}{{\tt
  1602.03560}}].

\bibitem{Dowdall:2019bea}
R.~J. Dowdall, C.~T.~H. Davies, R.~R. Horgan, G.~P. Lepage, C.~J. Monahan,
  J.~Shigemitsu et~al., \emph{{Neutral B-meson mixing from full lattice QCD at
  the physical point}},
  \href{http://dx.doi.org/10.1103/PhysRevD.100.094508}{\emph{Phys. Rev. D} {\bf
  100} (2019) 094508}, [\href{http://arxiv.org/abs/1907.01025}{{\tt
  1907.01025}}].

\bibitem{Dawson:2024ozw}
S.~Dawson, M.~Forslund and M.~Schnubel, \emph{{SMEFT matching to Z' models at
  dimension eight}},
  \href{http://dx.doi.org/10.1103/PhysRevD.110.015002}{\emph{Phys. Rev. D} {\bf
  110} (2024) 015002}, [\href{http://arxiv.org/abs/2404.01375}{{\tt
  2404.01375}}].

\bibitem{Boughezal:2022nof}
R.~Boughezal, Y.~Huang and F.~Petriello, \emph{{Exploring the SMEFT at
  dimension eight with Drell-Yan transverse momentum measurements}},
  \href{http://dx.doi.org/10.1103/PhysRevD.106.036020}{\emph{Phys. Rev. D} {\bf
  106} (2022) 036020}, [\href{http://arxiv.org/abs/2207.01703}{{\tt
  2207.01703}}].

\bibitem{Corbett:2023qtg}
T.~Corbett, J.~Desai, O.~J.~P. \'Eboli, M.~C. Gonzalez-Garcia, M.~Martines and
  P.~Reimitz, \emph{{Impact of dimension-eight SMEFT operators in the
  electroweak precision observables and triple gauge couplings analysis in
  universal SMEFT}},
  \href{http://dx.doi.org/10.1103/PhysRevD.107.115013}{\emph{Phys. Rev. D} {\bf
  107} (2023) 115013}, [\href{http://arxiv.org/abs/2304.03305}{{\tt
  2304.03305}}].

\bibitem{Aebischer:2022aze}
J.~Aebischer and M.~Pesut, \emph{{One-loop Fierz transformations}},
  \href{http://dx.doi.org/10.1007/JHEP10(2022)090}{\emph{JHEP} {\bf 10} (2022)
  090}, [\href{http://arxiv.org/abs/2208.10513}{{\tt 2208.10513}}].

\bibitem{Fuentes-Martin:2022vvu}
J.~Fuentes-Mart\'\i{}n, M.~K\"onig, J.~Pag\`es, A.~E. Thomsen and F.~Wilsch,
  \emph{{Evanescent operators in one-loop matching computations}},
  \href{http://dx.doi.org/10.1007/JHEP02(2023)031}{\emph{JHEP} {\bf 02} (2023)
  031}, [\href{http://arxiv.org/abs/2211.09144}{{\tt 2211.09144}}].

\bibitem{Alonso:2013hga}
R.~Alonso, E.~E. Jenkins, A.~V. Manohar and M.~Trott, \emph{{Renormalization
  Group Evolution of the Standard Model Dimension Six Operators III: Gauge
  Coupling Dependence and Phenomenology}},
  \href{http://dx.doi.org/10.1007/JHEP04(2014)159}{\emph{JHEP} {\bf 04} (2014)
  159}, [\href{http://arxiv.org/abs/1312.2014}{{\tt 1312.2014}}].

\end{thebibliography}\endgroup
\bibliographystyle{JHEP}

\end{document}